# Nondestructive, quantitative viability analysis of 3D tissue cultures using machine learning image segmentation

Kylie J. Trettner[1,2], Jeremy Hsieh[3], Weikun Xiao[2,†], Jerry S.H. Lee[1,2,4], Andrea M. Armani[1,2*]

[1] Mork Family Department of Chemical Engineering and Materials Science, Viterbi School of Engineering, University of Southern California, Los Angeles, CA, USA

[2] Ellison Institute of Technology, Los Angeles, CA, USA

[3] Pasadena Polytechnic High School, Pasadena, CA, USA

[4] Department of Medicine, Keck School of Medicine, University of Southern California, Los Angeles, CA, USA

[†] Now located at Appia Bio, Los Angeles, CA, USA

* Andrea M. Armani

Email: aarmani@eit.org

**Abstract**

Ascertaining the collective viability of cells in different cell culture conditions has typically relied on averaging colorimetric indicators and is often reported out in simple binary readouts. Recent research has combined viability assessment techniques with image-based deep-learning models to automate the characterization of cellular properties. However, further development of viability measurements to assess the continuity of possible cellular states and responses to perturbation across cell culture conditions is needed. In this work, we demonstrate an image processing algorithm for quantifying features associated with cellular viability in 3D cultures without the need for assay-based indicators. We show that our algorithm performs similarly to a pair of human experts in whole-well images over a range of days and culture matrix compositions. To demonstrate potential utility, we perform a longitudinal study investigating the impact of a known therapeutic on pancreatic cancer spheroids. Using images taken with a high content imaging system, the algorithm successfully tracks viability at the individual spheroid and whole-well level. The method we propose reduces analysis time by 97% in comparison to the experts. Because the method is independent of the microscope or imaging system used, this approach lays the foundation for accelerating progress in and for improving the robustness and reproducibility of 3D culture analysis across biological and clinical research.

## I. Introduction

Cellular viability is a fundamental metric used to characterize the growth characteristics and proliferative capability of cell or tissue cultures. Viability assays are used throughout biology and preclinical toxicology to understand a wide range of behaviors, such as changes in cell growth induced by therapeutic perturbations or change in cell culture conditions. Assaying techniques range from colorimetric indicator dyes[1–4] or fluorescence probes[5,6] to an analysis of the metabolic activity by chemically lysing the cells to determine the amount of ATP present using luminescence[7]. Results, typically reported from 0% to 100%, are obtained by evaluating the ratio of healthy to dead cells or the metabolic activity of healthy cells in a sample population. In some cases, cell viability assay results are supported by performing orthogonal measurements, such as sample imaging.

As the application of deep learning to image analysis has increased in the biomedical field[8,9], many researchers have found value in automating image analysis pipelines to extract quantitative information about the cellular systems studied[10,11]. These computational models can segment and identify single cells and cell types[12–14], predict phenotypes of the cells[15,16], assign fluorescent markers to cell images[17,18], and quantify viability of the cells within a given image[19]. The principles behind these models have enabled advances beyond single cell and apply to tissue-level analyses[20], medical imaging[21,22], and predictive diagnostics for diseases such as cancer[23–26]. However, a weakness of both the original viability assays and the imaging techniques is that they were initially developed for use in 2D culturing conditions in flasks/dishes or multi-well plate formats.

Recent research has demonstrated that cells grown in 3D culturing systems better recapitulate the matrix interactions and cell phenotypes found in physiological tissue[27]. The representation of these complex interactions provides an improved in vitro method for compound screening over 2D systems. While improvements in imaging technology have advanced the ability to capture qualitative information about 3D cell models and tissues[28–30], the need for longitudinal quantitative assessment of culture viability is still present. Various strategies are being pursued to address this challenge. One manufacturer has adapted their assay to create a 3D culture specific product. This product is called CellTiter-Glo 3D (CTG) and has quickly become an industry standard assay used to report viability by measuring the metabolic activity of the culture. However, the CTG development procedure requires complete cell lysis. For other methods, researchers have investigated the comparative use of previously developed assays[4] and recommend complementary imaging for experiments utilizing 2D specific reagents to cross reference results[31]. Although the field has begun adapting viability assays[32–34] and deep learning techniques to the dynamic nature of 3D systems[35,36], quantification of viability largely requires additional experimental steps for fluorescently labeling the culture system[37,38] or developing metabolic assays[39,40]. This remaining challenge provides a unique opportunity to utilize deep learning to augment 3D viability assays.

In the present work, we develop and validate a Segmentation Algorithm to Assess the ViabilitY (SAAVY) of 3D cultures. We designed SAAVY to automatically identify and analyze features common to spheroid and organoid structure that experts correlate with cellular viability, such as the transparency of the spheroid and the overall morphology[41–43]. SAAVY is designed for use with label-free optical images that are saved in the universal imaging formats (png, tiff, jpeg), making it independent of the microscope system used to acquire the images. We trained and tested SAAVY against clear and noisy backgrounds. This approach ensured that SAAVY can withstand a degree of background noise that can arise from common biological defects, such as dead cell fragments or matrix particulate deposits. SAAVY calculates the viability of each uniquely identified spheroid in an image and an overall average viability for all 3D structures present in a well. It also reports total spheroid count per image, spheroid radius, spheroid area, and other metrics of relevance. The total analysis time per well is approximately 0.3 seconds. This type of integrated analysis provides insight into a biological system's response at both the individual spheroid and entire well level. Lastly, SAAVY is agnostic of microscope system or manufacturer and does not require fluorescent or colorimetric indicators, enabling longitudinal response studies to be performed.

The accuracy of SAAVY in analyzing pancreatic ductal adenocarcinoma (PDAC) spheroids in clear and noisy backgrounds was validated through a blinded comparison with a pair of spheroid analysis experts. Subsequently, a series of application-driven experiments were conducted. We first compared SAAVY analysis with an industry standard metabolic assay for 3D cultures, including spheroid expert analysis as ground truth. To challenge SAAVY's ability to detect viability changes in response to a perturbation, we then performed a label-free imaging-based longitudinal study investigating the effect of an FDA-approved therapeutic on the pancreatic cancer spheroids.

## II. Results and Discussion

To segment the biological regions of interest in an image, prior image recognition approaches focused on utilizing relatively simple edge detection techniques to identify spheroid boundaries[44,45]. However, edge detection with watershed has poor reliability when spheroids or organoids overlap or when background noise is present, limiting the utility to images with low spheroid density or complete separation of cell colonies[40]. Other characterization workflows reduced or eliminated this overlap by altering culturing conditions (e.g. growing singe spheroids per well) or image acquisition settings (e.g. increased magnification on the sample wells to obtain one spheroid per image) on a case-by-case basis[40,44,46]. This bespoke approach relies on image-stitching and partial data rejection of overlapping regions of interest, which can lead to unintentional bias in the final data set[47,48]. Our goal is to develop an algorithm that can analyze an image of an entire well with minimal data rejection while resisting the influence of background noise, which requires a different approach to segmentation.

As a proof of concept, PDAC spheroid samples are used. PDAC spheroids are of a cystic phenotype and tend to grow randomly throughout the well[49]. This growth pattern leads to a high frequency of overlapping cell structures and makes them difficult to analyze using edge detection machine-learning approaches. Healthy cystic spheroids are distinguished by their open lumen and transparent centers with distinct, circular edges when viewed in plane. The transparency and circular morphology of healthy spheroids are in stark contrast with the opaque, blebbed spheroids that characterize dead spheroids of this type. This correlation between transparency and morphology and spheroid viability in brightfield images has been previously studied[50,51]. For example, this characteristic indicator is seen in kidney[52], ectocervical[53], colon/intestinal[54,55], nasal epithelial[56] and liver[57] model systems. As part of this work, an informal survey of spheroid and organoid experts across cancer fields was performed, and they identified the same metrics. These results are included in the supplementary information. As a result, the ability to correlate opacity to spheroid viability forms the foundation of our label-free image-based quantification approach.

### A. SAAVY design

An overview of SAAVY is presented in Figure 1 and details are included in the SI. SAAVY is first created by fine-tuning a pre-trained Mask R-CNN model in PyTorch[58]. This implementation is particularly attractive due to its improved ability to separate overlapping features. This transfer learning approach reduced the total cost for bespoke images of cystic spheroids and allowed us to move forward with a relatively small number of expert-annotated images[59]. We further refined our model with a balanced dataset consisting of 24 images of PDAC spheroids with clear and noisy backgrounds that spanned the entire viability range. Details on this process are contained in the Supplementary Information.

SAAVY analyzes brightfield, label-free images of tissue culture spheroids by segmenting each individual spheroid and outlining the identified region on the output image. Subsequently, SAAVY quantifies the viability of each uniquely identified spheroid using the average intensity of the segmented region compared to the background using a weighted average model as detailed in the Supplementary Information. This approach allows for an increased level of background noise to be present without negatively impacting the assessment ability. Several other metrics are calculated by SAAVY including the average viability, the average spheroid size and total spheroid count across each image. With our hardware configuration, each image required 0.3 seconds for the entire analysis process. While viability assessments of 3D cultures are routinely performed as a quality check during experimentation by experts, the single-spheroid level quantitative analysis across an entire well is not easily obtained from expert analysis and is not possible using biochemical measurements. Therefore, SAAVY provides several orthogonal dimensions for analytics on an accelerated timescale.

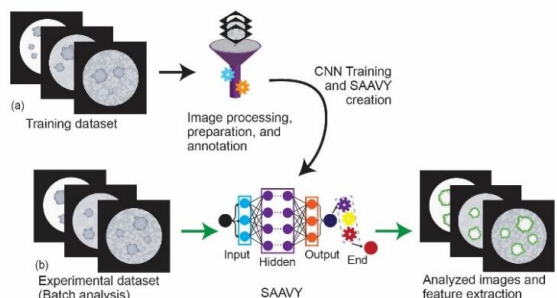

Figure 1 - Overview of the SAAVY Training and Validation approach. A) The training method for SAAVY is identified where we processed the images saved in png file format for annotation and user-supervised training. The viability algorithm is developed during this step based on expert-determined characteristics for generating viability estimates. However, to avoid overfitting, the spheroid experts were not directly involved in the training. The training data set was 30 images and required approximately 1 hour.  B) Batch analysis was run on the experimental (validation) data set images were passed through SAAVY, which identified the mask of the region of interest in green and output a .csv file with other measured information about the image. Analysis time per image was 0.3 seconds. Both training time and analysis time are dependent on the computational power available and could be further accelerated.

The initial training and validation data set contains a series of label-free microscopy images of PDAC spheroids cultured either in a clear or a noisy culture matrix. The images were annotated by the programming expert with no input from either spheroid expert. The noisy matrix was created by intercalating opaque nanoparticles throughout the gel. Images are acquired at days 0, 4, and 6 of spheroid growth at 4x magnification, and each image captures the entire 10 µL gel seeded in a 96-well plate. The resultant data set contains 1,328 widefield images taken using an ECHO Revolve microscope. Notably, no images or wells were rejected from the data set. Additional experimental details are included in the SI.

Representative examples of SAAVY-analyzed images across all days and both clear and noisy backgrounds are presented in Figure 2. Both the original and the analyzed images are shown. As can be seen, SAAVY is able to identify spheroids in the presence of potential confounds and in cases where spheroids are overlapping. Notably, the entire image was analyzed at once. As part of this work, SAAVY detected and provided information about 114,726 unique spheroids across the 2,792 images taken on either an ECHO Revolve or an Operetta CLS microscope. The compatibility with two different imaging systems demonstrates its potential impact on the field.

The overall predictive accuracy of SAAVY was determined by calculating the intersection over union (IoU) for a subset of images randomly selected from the dataset highlighted in Figure 2 for the ECHO Revolve and from a different data set for the Operetta. The average IoU for the ECHO is 0.622, and the average IoU for the Operetta is 0.653. Details on this method and additional analysis are included in the SI.

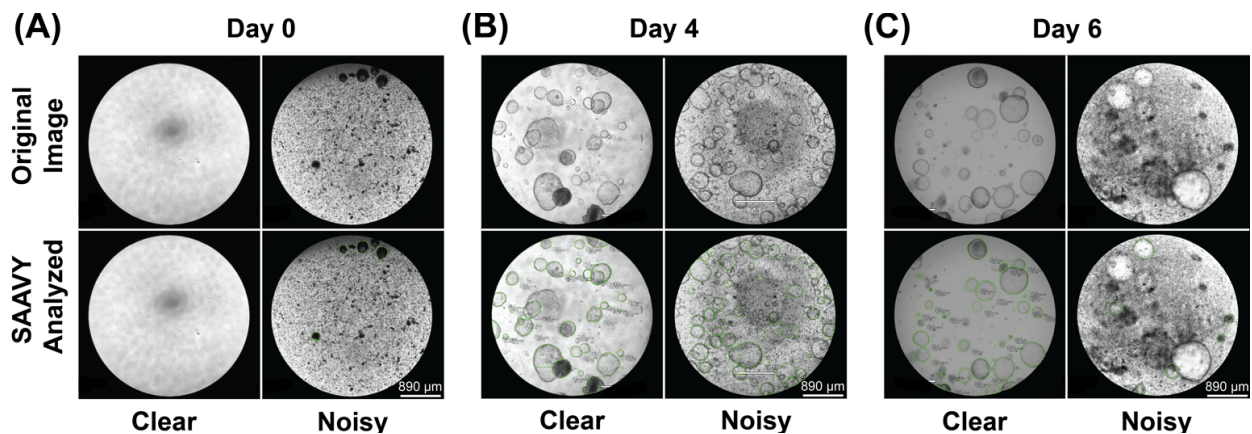

Figure 2 - Images representative of before (top row) and after (bottom row) SAAVY analysis. The images are further subset by day: a) D0, b) D4, c) D6). For each day, both clear (no added nanoparticles) and noisy (5 mg/mL of added nanoparticles) images are shown. These are representative images selected from the total 416 images analyzed. Scale bar is 890 µm for all images.

## B. SAAVY to Expert comparison

SAAVY performance was evaluated as compared to a pair of blinded spheroid experts for the entire image population of clear and noisy background gels. The justification for expert selection and the blinding method is detailed in the SI. Additionally, before comparing the expert assessments, their findings were harmonized. The steps involved in this process are detailed in the supplementary information.

Two indicators of SAAVY performance are evaluated: spheroid/no spheroid identification (ID) and live/dead detection (LD). ID is characterized as SAAVY assigning the appropriate value to each image according to the absence (value of -1) or presence (values of zero and above) of spheroids in the image. LD is characterized as SAAVY assigning viability values that correlate to live (>0%) or dead (0%) spheroids. This corresponds to the typical binary interpretation of other viability assays to keep analysis consistent with previous research. On Days 4 and 6, ground truth was determined by Expert 1, and the details for this decision are included in the Supplemental. Both ID and LD performance are quantified using the F1-score and summarized in Table 1.

The quantification of differences between the distributions was first calculated to compare SAAVY to each expert. For this quantification, Earth Mover's Distance (EMD) was used[60]. Values of 0 correspond to less distance between the distributions, or classically that the effort is minimized to transform one distribution into the other. The EMD values are represented by the 'Similarity' row in Table 1. Further, the reliability was quantified using Krippendorff's alpha, a measure of inter-rater reliability. Values close to 1 suggest perfect reliability whereas values closer to zero and negative values suggest poor reliability and systemic differences, respectively. The supplementary material contains a comparison of other common metrics and data correlations that we calculated for this analysis.

Table 1 - Summary of calculated metrics comparing spheroid Expert 1, spheroid Expert 2, and SAAVY throughout each sample sub-grouping (day and background type). Spheroid identification (ID) and live/dead (LD) analysis across all days and background types are presented first. Dashes (-) are noted in columns where that method was used as either the ground truth (Expert 1) or standard for comparison (SAAVY). Similarity and reliability are then quantified for each distribution comparison. We used the Earth Mover's distance to quantify the similarity between the two distributions and Krippendorff's alpha to quantify the reliability of each expert. Equations for accuracy and F1-score calculations are included in the Supplemental. Red cell highlights indicate better performance whereas blue values indicate weak performance.

|  | Day 4 | | | | | |
|---|---|---|---|---|---|---|
|  | Clear | | | Noisy | | |
|  | Expert 1 | Expert 2 | SAAVY | Expert 1 | Expert 2 | SAAVY |
| ID F1-Score | - | 1 | 1 | - | 0.992 | 0.992 |
| LD F1-Score | - | 1 | 1 | - | 0.992 | 0.992 |
| Similarity | 0.029 | 0.073 | - | 0.021 | 0.036 | - |
| Reliability | 0.531 | -0.202 | - | 0.869 | 0.735 | - |

|  | Day 6 | | | | | |
|---|---|---|---|---|---|---|
|  | Clear | | | Noisy | | |
|  | Expert 1 | Expert 2 | SAAVY | Expert 1 | Expert 2 | SAAVY |
| ID F1-Score | - | 1 | 1 | - | 0.984 | 0.992 |
| LD F1-Score | - | 0.914 | 0.891 | - | 0.962 | 0.954 |
| Similarity | 0.045 | 0.048 | - | 0.016 | 0.011 | - |
| Reliability | 0.869 | 0.861 | - | 0.863 | 0.854 | - |

On Day 4, SAAVY shows a decrease in reliability compared to Expert 2 in clear samples (Table 1). The same is not true for SAAVY as compared to Expert 1 in clear backgrounds, and the reliability is improved for both experts for noisy background gels. The similarity also follows this trend for Day 4. On Day 6, SAAVY is nearly equally reliable when comparing to both experts for both background types. However, the SAAVY distribution is more like the spheroid experts for noisy backgrounds as compared to clear backgrounds. The distributions are visualized for all permutations in the SI for further comparison.

Day 4 showed clear improvement for both experts and SAAVY with F1-scores of 1.0 for both ID and LD analyses, which suggests perfect agreement with the ground truth. When spheroids are observed after the given time for growth, they are distinguishable from the background. The fact that SAAVY matches experts at this task suggests that the algorithm can successfully identify and analyze spheroids in clear backgrounds comparably well to human experts. LD analysis matched this trend of perfect F1-scores (1.0) for both SAAVY and Expert on Day 4. For noisy backgrounds, the same holds true though the F1-score decreases slightly to 0.992. Day 6 is when differences between SAAVY performance and the experts begin to appear. Notably, SAAVY outperforms Expert 2 as compared to ground truth for spheroid identification. For LD analysis on clear backgrounds, it is 0.91 for Expert 2 compared to 0.89 for SAAVY. On noisy backgrounds, this is further improved to 0.96 for Expert 2 and 0.95 for SAAVY.

Although SAAVY has slightly lower F1 scores than Expert 2 for live/dead performance, this difference is likely an artifact related to the size of detectable viability differences characteristic to each expert. SAAVY assigned viability values to a one-tenth of a percent as a direct scaling of the viability per pixel is 0.4%. Additionally, as can be seen in Figure S5, SAAVY did not exhibit a bias in assignments towards any viability range. In contrast, after data harmonization, the spheroid experts classified viability in 10% increments. As a result, any spheroid with a viability below 10% is classified as dead. Additionally, the distribution of viability scores was biased towards the extreme ends. Therefore, SAAVY may classify an image of spheroids as alive although they are noted dead by experts due to its improved incrementation and non-biased assessment. To test this hypothesis, a ROC threshold analysis was completed to evaluate the level at which the threshold for live or dead can be raised to while optimizing the F1-score for SAAVY in clear samples. The threshold for SAAVY can be raised to 2.43% for clear backgrounds leading to an F1-score of 0.91, which matches Expert viability prediction while maintaining improved resolution.

**C. Comparison against industry standard viability assay**

CellTiter-Glo (CTG) is a metabolic viability assay that has been adapted and optimized for use when analyzing 3D tissue culture systems[39,40]. Unlike image analysis methods, CTG relies on correlating the amount of cellular ATP released from lysed cells to the intensity of the luminescent signal generated by an enzymatic reaction. These assays are evaluated on a well-by-well basis through normalization of the well signal to positive and background control measurements. Although this assay is accepted and widely used as the industry-standard, previous research utilizing this method has highlighted several limitations. For example, specific metabolic effects may be obscured when measuring viability on its own[61–63]. Inconsistent experimental procedures are also of concern for CTG measurements. Previous studies suggest measurements are dependent on variables such as the shaking time and wait time for equilibrium before measuring the plate luminescence[39]. This variation may lead to errors in the normalization process standard for reporting the results of this assay. Image-based methods remove the question of experimental variation on reported viability.

We compare SAAVY viability to CTG viability measurements through a non-inferiority study to test if SAAVY performs no worse than the CTG method at characterizing viability. PDAC samples grown in either clear or noisy matrices were analyzed by the experts, CTG, and SAAVY. This approach resulted in 136 images with matched CTG data. Without noise in the matrix, the CTG viability reading exceeded 100% a total of seventeen times (24.3% of the measurements). Once matrix impurities were introduced, the CTG viability reading exceeded 100% in 39 samples (57.4% of the measurements). It should be noted that prior research has shown an increase in metabolic activity when grown in similar nanoparticle matrices[64], so the CTG results agree with prior findings.

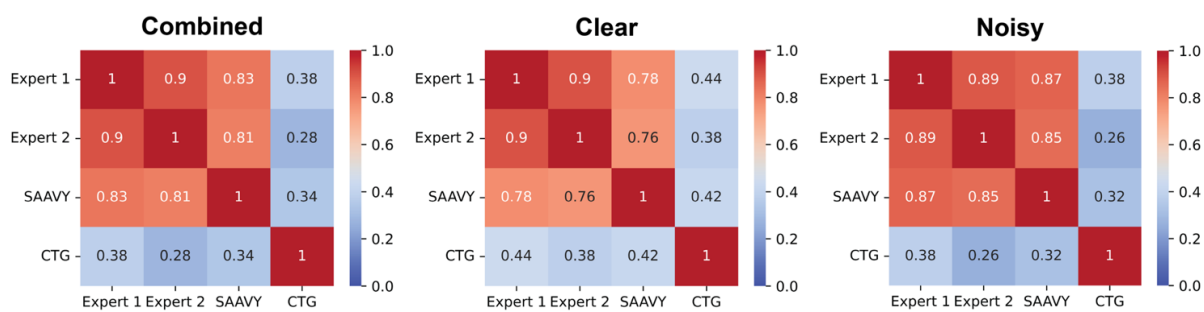

Figure 3 - Pearson's R correlation analysis presented as heatmaps comparing Experts, SAAVY, and CTG for the Overall (Combined), Clear background, and Noisy background datasets. Red indicates R values closer to 1 (perfect, positive correlation) and blue indicates no correlation. CTG values are poorly correlated to all other measurements regardless of the data subset. This is likely due to the relative interpretability of CTG results.

We compared CTG to SAAVY using Pearson's correlation, Earth Mover's distance for similarity, and Krippendorff's alpha for rater reliability across Day 6 data for clear and noisy backgrounds. Figure 3 summarizes the results of the correlation comparisons. SAAVY and CTG have similar distributions for clear and noisy background types with EMD values of 0.023 and 0.010, respectively. We believe this agreement is due to the absence of spheroids in some images in this dataset, as the noisy background either obscured detection or killed the growing spheroids by this day. Figure S10 includes a histogram breakdown of this data to visualize the distributions. The reliability analysis, however, suggests that CTG is a more reliable rater on clear gels (k-alpha, 0.384) compared to noisy (k-alpha 0.076).

We could not complete a confusion matrix analysis for SAAVY and CTG because CTG cannot produce a negative result. We perform a threshold analysis on CTG to determine what the viability percent could be raised to with the goal of enhancing the overall performance of the CTG test. However, because of the inability of CTG to provide negative (0% viable by LD analysis) measurements, we looked to a metric other than F1-score during our threshold analysis. Youden's J-statistic was used to determine that the cutoff for CTG can be raised to 18.00% and 48.15% for clear and noisy backgrounds, respectively. This suggests that the CTG assay is not specific when capturing cell state characteristics and points toward the potential impact of an image-based companion for viability measurement.

**D. Non-destructive tracking of perturbation longitudinal impact on 3D cultures**

One current hurdle in longitudinal studies using high-throughput, automated 3D culture methods is the requirement of labeling the sample to assess viability. A non-destructive, label-free approach would allow continuous monitoring of the same spheroid, increasing research rigor and allow primary cells and samples to be used. As a step in this direction, SAAVY's ability to longitudinally track PDAC spheroid growth within the same sample is demonstrated by analyzing label-free images and creating a viability response curve.

An Operetta CLS high content imager is used to automate the image acquisition process. During this experiment, we took z-stack images of each well in a 96-well plate 20 µm apart from 0 µm at the hydrogel/well-plate interface to 700 µm at the top of the hydrogel resulting in 36 images per each well. Imaging for this experiment was conducted on six days following the seeding of the experimental plate on day 0. The final longitudinal image set included 12,528 images where 2,376 images were selected for analysis (105,723 spheroids). The z-height used in Figure 4a-c was located at our chosen mid-plane in the well at 320 µm. Five planes above and below the midplane were included for individual spheroid analyses. However, to scan the largest region within the center of the gel to test uniformity of SAAVY analyses we chose planes 40 μm apart. Further details are included in the SI.

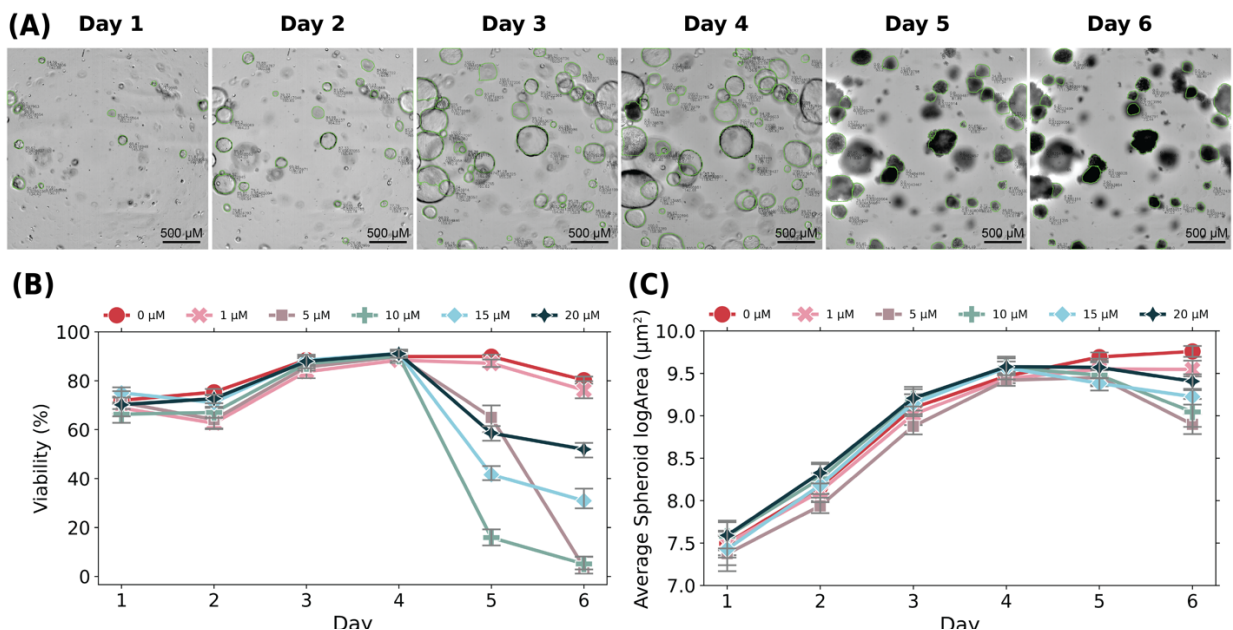

Figure 4 - Longitudinal analysis of PDAC response to gemcitabine evaluated at whole-well resolution. A) Images on the top row depict a representative sample well during each day of a six-day dose-response assay. The chosen well was perturbed with 10 μM drug (gemcitabine) on Day 4 after imaging. Images are of the same focal plane (320 μm) throughout the course of the experiment. The scale bar is 500 μm. B)/C) Line plots of SAAVY application to analyze the viability and growth of spheroids at whole-well resolution across the dose-response experiment where: B) the raw viability of each drug-response group over the assay and C) the average log-transformed spheroid area (μm$^2$) of each drug-response group over time.

Gemcitabine, an anti-metabolite therapeutic commonly used to treat pancreatic cancer, was used for compound perturbation at concentrations of 1, 5, 10, 15, and 20 μM. We included wells of untreated PDAC spheroids for a positive control. Treatment was added after imaging on day 4 to all wells using appropriate volumes of a stock 100 μM solution of gemcitabine in DMSO. An example well from the 10 μM perturbation group was highlighted in Figure 4a to note the change in morphology and color of spheroids over time in this experiment.

Whole-well viability, plotted by treatment group, is presented in Figure 4b. Critically, we did not normalize the reported viabilities in this plot and used the raw output from SAAVY to emphasize the low error within repeated sample conditions. The decreased viabilities reported by SAAVY on Days 1 and 2 is likely due to the smaller size of the spheroids at this time, specifically taking into consideration that our training was done on Day 4 and Day 6 spheroids. However, the stability of viability across Days 3 and 4 of spheroid growth is suggestive of the overall reliability of SAAVY to track samples across different days.

SAAVY detects viability responses across all sample groups. The untreated and 1 μM groups trend similarly for viability across all days. Interestingly, the mechanisms of compound diffusion are potentially identified in our image-based assay as well. We note a delayed spheroid-death response from the 5 μM concentration, perhaps due to diffusion time through the gel leading to decreased availability of compound 24-hours after treatment compared to 48-hours post-treat. In contrast, we see a smaller change in viability across the 15 and 20 μM concentrations. Upon qualitative analysis of the images, we observed a distinctly different morphology of the spheroids. Spheroids at these higher compound concentrations look like they have collapsed on themselves, compared to the lower (5 and 10 μM) where the spheroid looks like it has exploded and has characteristic blebbed edges suggesting cell apoptosis (as shown in the day 5 and day 6 images in Figure 4a).

We investigate if the decreased viability trends with spheroid area, as we expect the area to decrease as the spheroids die. This is confirmed by visualizing the average logarithmic transformed area across each day of the assay, seen in Figure 4C. We see an upward trend in spheroid growth across all treatment groups through Day 4 of the assay. This metric allows us to investigate further trends in spheroid response to compound treatments. For the control group, we see continued growth through the final day of imaging. We see an expected decrease in spheroid size from Day 4 to Day 6 in the 5, 10, 15 and 20 μM treatment groups that correlates with the decreased viability seen on these days. For the lowest concentrations of treatment, the size increases until Day 5, though at a slightly lesser rate than the control group, where it then stays constant between Day 5 and Day 6. When taken into context with the viability data, this suggests that this smaller concentration of compound may be enough to stall continued growth without eradicating the tumor cells. A potential rate-based result underscores the important role that image-based surveillance methods play during cell viability experiments.

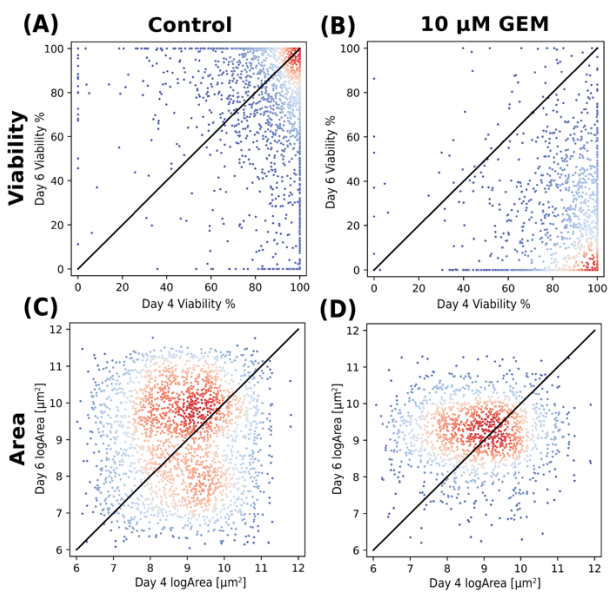

Figure 5 – Scatter plots of individual spheroids from Day 6 plotted against Day 4 spheroids colored according to a gaussian kernel density estimation where blue is lower density and red is higher density. The plots either present the viability (A and B) or the log(Area) (C and D) for either the control group or the 10 μM gemcitabine (GEM) concentration perturbation group.

To further investigate these rate-based changes, we performed a single spheroid level analysis, and we plot day 6 metrics against day 4 metrics on spheroids taken from the same sample. The scatter plot representation in Figure 5 underscores the visual changes that take place in the sample with the introduction of perturbation compound. The black diagonal line in each plot is a guide to the eye. Any spheroid above the line is increasing in viability or area from day 4 to day 6; conversely, data below the line has experienced a decrease.

In the viability plots (Figure 5a and 5b), we see the highest density of spheroids, represented by the red colored points on the scatter, moving from the top right of the plot in the control group to the bottom right of the plot in the 10 μM concentration group. This distribution shift clearly highlights the negative viability trend in individual spheroid response. The area plots in Figure 5c and 5d show an interesting nuance to the spheroid size. Where the whole-well analysis suggests that the area of spheroids decreases with perturbations and cell death, additional information is revealed when the data is analyzed at the single spheroid level.

Specifically, in the control group, two spheroid size populations are clearly identifiable. The larger of the two is increasing in size, but a small population is decreasing. In contrast, in the GEM-treated samples, there is a uniform size population that is stagnant in size. When analyzed in conjunction with the data in Figure 5b, one possible conclusion is that these spheroids are all dead. This type of analysis that blends population-level and single spheroid level data is not possible with other approaches and opens the door to reveal new mechanistic insights.

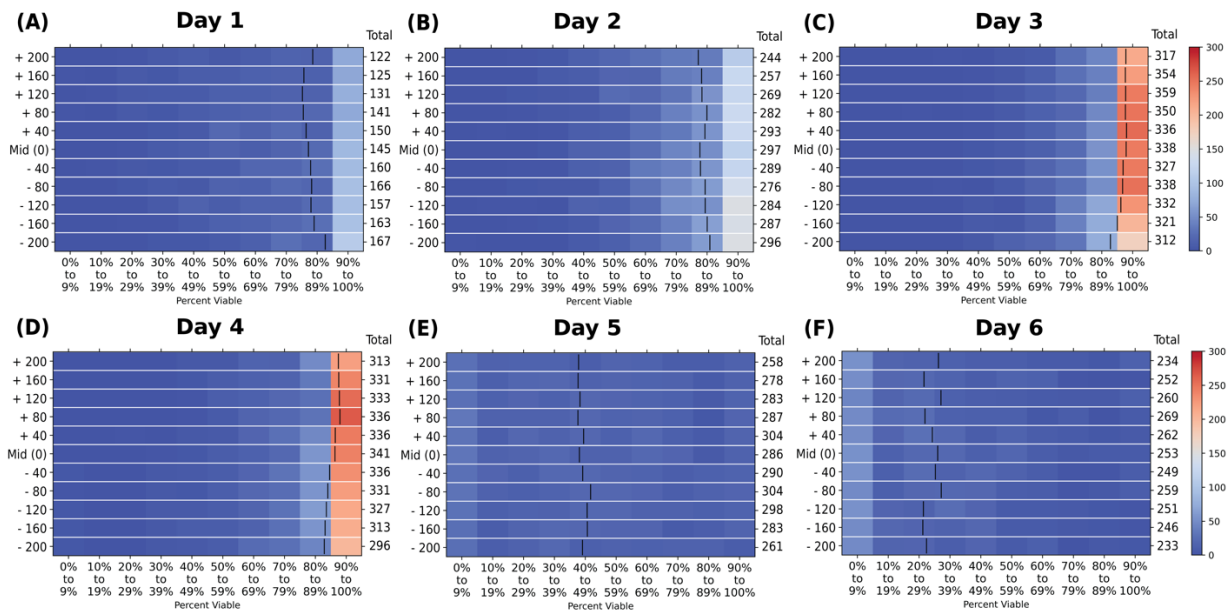

Figure 6 – 2D histograms of spheroids in the 10 µM perturbation group separated by the focal plane from imaging against the percent viability (binned in 10% increments) of all individually identified spheroids in the data set. The total number of spheroids identified per each plane are noted in the "total" column between the plot and the color bar. The black bar within each plane grouping is the group's average viability. The color bar represents the total number of spheroids within each viability bin, where red colors indicate high counts (closer to 300) and blue indicate low counts (closer to 0) of spheroids within each of the viability bins. Data are segmented by day accordingly. Subplots are for the day of experiment: A) Day 1, B) Day 2, C) Day 3, D) Day 4, E) Day 5, F) Day 6.

To round out the analysis, we compare images at different z-planes within the well and present this data as 2D histograms separated by day in Figure 6. We analyzed 5 image planes above and below the mid-plane, which was the data presented and analyzed in Figure 4. Each subsequent plane was 40 µm separated from the one before it to span the middle region of the spheroid-laden hydrogel. This spacing avoids interface effects, allowing us to capture the bulk characteristics. Further details are included in the Supplemental Information.

We extracted information on each individual spheroid within the images corresponding to the plane. All the 105,723 segmented spheroids are present in the plots detailing this analysis, with the other perturbation groups and area visualizations included in the Supplemental Information. In Figure 6, we see the same trend of increasing spheroid viability across Days 1-4 until perturbation followed by decreasing viability on Days 5 and 6. Importantly, we note the consistency of SAAVY to assign variabilities within each of these planes. The average for each plane distribution is consistent across each grouping and each day.

The individual spheroid analysis presented in Figures 5 and 6 suggest that SAAVY can uniformly identify the overall spheroid population within the growth matrix, and SAAVY analysis does not lose sight of overall response dynamics of spheroids within the entire culture.

## III. Conclusions

We develop and validate a method to quantitate viability of 3D tissue cultures in a label-free, non-destructive, and longitudinal manner that easily integrates into existing tissue culture imaging procedures. We showed comparable quantification using our algorithm, SAAVY, to human expert estimation in both clear and noisy background hydrogels with high tolerance to noise. We found improvements in classification ability and sensitivity over standard CTG assay measurement. Further, we showed the ability of SAAVY to longitudinally track 3D spheroid growth and viability for both untreated and therapeutically treated spheroids. Our algorithm uses the same principle that human experts use to evaluate images yet streamlines the process by automation. This facilitates integration of our viability method into standard tissue culture procedures in a robust and time-independent manner. The ability of our algorithm to output results of "no spheroid" as well as quantitative morphological metrics when analyzing 3D tissue culture constructs adds another analysis level on top of typical live/dead classification and can provide more information to an experimentalist regarding growth characteristics of a sample.

With the augmentation of biomedical imaging with computer vision and other CNN approaches, it is possible to further develop label-free imaging methods. Our system only covered the supervised training of one tissue spheroid type, PDAC. The open-source nature of our algorithm may allow for user-specified training of different kinds of 3D spheroid culture images as well as tunability of the viability algorithm to suit the characteristics of other cultures. There is opportunity to expand the training data of SAAVY beyond one-spheroid type and utilize other deep-learning methods, such as unsupervised learning, to expand the capabilities of SAAVY. This work considered the plane of best representation of each imaged spheroid in the experimental plates. To truly evaluate efficacy of compounds and materials, it is important to utilize the full reconstructive capabilities when merging 3D imaging with deep learning. The results of this study and other research[65] suggest the importance of studying 3D cultures to support drug discovery through dynamic analyses. In this context, improved monitoring may improve rigor in its ability to compare across experiments. The time-agnostic and non-destructive manner of this algorithm quantifies the qualities that researchers often judge when monitoring their culture. With further expansion of our training dataset to include various disease types and 3D tissue culture systems, we believe that SAAVY may prove a useful tool for real-time analysis and can complement image-based 3D culture assays where viability must be assured.

## IV. Methods
### A. Tissue culture

Mouse-derived pancreatic cancer (PDAC) spheroids (line 8-14F-7: KRAS G12D, PTEN loss, COX2 overexpression, female, 2 weeks old at the time of sacrifice). Pancreatic cancer spheroids were cultured using an established protocol[66] included in the Supplemental. For samples that do not include any nanoparticles, the cell-laden hydrogel solution was plated in 10 µL increments in the center of the wells on an opaque-walled 96-well plate (Corning). For samples that included nanoparticles, nanoparticle solutions were combined with cell-laden solutions to the desired concentration of cells and nanoparticles then plated in the same manner mentioned above.

Images were taken of all seeded wells on the day of seeding (Day 0, D0). The plates were left to incubate until day 4 (D4) where negative control wells were treated with 10 µM gemcitabine added to the cell media and images were captured of all wells. We incubated the plates for an additional two days until day 6 (D6) where we took final images of all then performed a 3D CellTiter Glo (CTG) assay (Promega) to measure the metabolic activity of the spheroids grown in each condition. CTG was performed according to the documented protocol provided by Promega and read using a BioTex plate reader for luminescent detection at 560nm. Dose-response assays were plated in the same manner described above and imaged at 320 μm above the bottom of the well plate for consistent cross sections across all six days of experimental growth. Images were exported from the Operetta CLS microscope with a brightfield correction factor applied to images to reduce vignetting from background light within the imaged gel.

**B. Image datasets**

Our datasets include brightfield images taken during various experiments, as noted in the tissue culture methods section above. PDAC images were taken on an ECHO Revolve with a 4x/0.13 objective, or Operetta CLS with 5x/0.16 objective lens. ECHO images were saved and exported in TIFF format. Images from CLS were exported as PNG with a brightfield correction applied by the instrument software to remove a vignette from the instrument's inhomogeneous light source. Both were converted to JPG and digitally resized to 1024x1024 for SAAVY viability analysis. The complete PDAC set included 1,328 images. Of these, 24 images were randomly selected for training, and 416 images were used for the experimental viability estimations where 136 were CTG matched. The longitudinal data set included 12,528 images where 2,376 were selected for analysis.

**C. Data training, pre-processing, and final run time**

We first train SAAVY using a pre-existing image set from MS COCO pretrained general model for transfer learning. We trained for 20 epochs. Based on analysis, we used the 15th checkpoint for image production due to continuous loss after this point (Figure S3). Using the training dataset detailed above, we annotate the 24 training images using VIA Image Annotator (2.0.11, Oxford) to specifically identify cell spheroids. Total computational time for SAAVY on the 416-image set is 2 minutes and 5 seconds (running an RTX 3080 and Intel Core i9-10850K stock).

**D. Post-hoc analysis**

We assess confusion matrices for two classifications: spheroid/no spheroid and live/dead detection. Accuracy and F1-score (SI, EQ-2,3) were calculated where appropriate to get an overview of the performance of SAAVY as compared to experts and how the experts compare to each other. We assessed data matching by first converting data to their respective probability density function and then calculating the multiple distance metrics to assess the distances between all viability estimation methods (Table S4).

Before performing any statistical analyses, we assessed overall normality of the data according to the group analyzed. Where appropriate, non-normal data was estimated normal according to the central limit theorem and the appropriate groupwise analyses were applied (either repeated measured analysis of variance with Tukey post hoc test or related t-test for normal data or Wilcoxon for nonparametric data from small samples). All analysis was completed in Python with the appropriate libraries noted in the SI.

**E. Supplementary Material**

We provide complementary analyses and expand details of the included analyses in this manuscript in an extensive supplementary file. The supplemental includes five distinct sections for 1) 3D culture and experimental preparation methods, 2) Dataset generation, 3) SAAVY development, 4) analysis, and 5) experimental metadata. We provide extensive data from our single spheroid and planar analysis in additional plots.

**V. Acknowledgments**

We would like to acknowledge and thank Reginald Hill, Ph.D., for PDAC spheroid culture access and training for use. Further, we extend our gratitude to Naim Matasci, Ph.D., the Computational Team and Cell Line Team of the Ellison Institute of Technology for their discussions regarding analysis methods, cell growth standardization, and best procedures throughout the experimental and analysis phases of this manuscript's preparation.

The authors would like to thank the Office of Naval Research (N00014-22-1-2466, N00014-21-1-2048), National Science Foundation (DBI-2222206), and the Ellison Institute of Technology for funding this work.

Disclaimer: The contents of this publication are the sole responsibility of the authors and do not necessarily reflect the views, opinions, or policies of the USUHS, the Henry M. Jackson Foundation for Advancement of Military Medicine, Inc., the Department of Defense, the Department of the Army, Navy, or Air Force. Mention of trade names, commercial products, or organization does not imply endorsement by the U.S. Government.

## VI. Data availability statement

The data that support the findings of this study are openly available in Zenodo at http://doi.org/10.5281/zenodo.10086367 and GitHub at https://github.com/armanilab/SAAVY[67].

## VII. Author Declarations

Competing Interest Statement: J.S.H.L. serves as Chief Science and Innovation Officer for Ellison Institute, LLC (paid); consultant for Henry M. Jackson Foundation (paid); scientific advisory board for AtlasXomics, Inc. and ATOM, Inc. (unpaid, travel support) outside the submitted work. A.M.A. serves as the Senior Director of Engineering and Physics for Ellison Institute, LLC (paid).

Author Contributions: K.J.T. performed particle synthesis, tissue culture experiments, imaging, and data analysis; K.J.T., J.H., and A.M.A conceptualized the SAAVY algorithm; J.H. created code, annotated, and trained data; K.J.T and W.X. performed expert analysis; K.J.T., J.H., and A.M.A. wrote the manuscript; W.X. and J.S.H.L edited and approved of the final manuscript documents.

# Nondestructive, quantitative viability analysis of 3D tissue cultures using machine learning image segmentation

Kylie J Trettner[1,2], Jeremy Hsieh[3], Weikun Xiao[2], Jerry SH Lee[1,2,4], Andrea M. Armani[1]*

*Email: aarmani@eit.org

**Table of Contents**

## 1 <u>3D Culture Preparation</u>

### 1.1 <u>Nanoparticle Synthesis</u>

We synthesized iron oxide ($Fe_3O_4$) nanoparticles using a previously reported air-free, coprecipitation method summarized here. First, we purged 20 mL of DI water in a three-neck round bottom flask with nitrogen while we measured 1 g of iron (II) chloride and 0.4 g of iron (III) chloride in an argon glovebox. We transferred the iron chloride reactants and added them to the reaction flask using a solids addition flask. We then stirred the reaction solution using a stir bar, submersed the reaction flask in a mineral oil bath, and heated it until the reaction solution reached 80 ºC. Once this temperature point was reached, 5 mL of ammonia hydroxide was added dropwise to the reaction flask. We increased the stirring speed during this step to avoid possible morphological changes to the nanoparticles caused by the magnetic stir bar. The reaction refluxed for one hour before the flask was removed from heat and allowed to cool. Once cool, we collected the magnetic particles into a falcon tube and centrifuged the collected material. After centrifugation, we decanted the water and then resuspended the particles in DI water to wash. We repeated the centrifugation and wash steps 3x before massing the final solutions to determine the yield and concentration of nanoparticle solutions. Using the nanoparticles, a series of solutions were prepared. The final concentrations of nanoparticles in the studied seeded wells were: 0.5, 1, 1.5, 2, 2.5, 3, 5, and 10 mg/mL.

## 1.2 Hydrogel Preparation

Two hydrogel materials were used during this study: 1) Matrigel (Corning) and 2) Basement Membrane Extract or BME (Cultrex). The specific Matrigel used was growth factor reduced, phenol red-free (Cat. No. 356231) of one of two lots (0258005 or 0322001). The BME used was reduced growth factor (Cat. No. BME001-05) of one of two lots (1617263 or 1661425). Matrigel and BME are prepared by first thawing the initial bottle (either 5 or 10 mL) on ice overnight in a 4ºC cold room. Once thawed, 1 mL aliquots were made with extra precaution not to create any bubbles when transferring. Aliquots were placed in -20ºC (Matrigel) or -80ºC (BME) freezer until use. When preparing for cell seeding, aliquots were placed on ice at room temperature in the tissue culture hood (biosafety cabinet) for up to an hour until thawed. We placed the 96-well plates in an incubator to warm up while the gel was thawing.

Once thawed, the hydrogel liquid is ready for use. We removed gel from the cell-laden or mixed nanoparticle-gel solutions in 10 µL increments and plated it in the middle of the well. We used 4-6 replicates of each condition to average out potential experimental variabilities. Once the plate was seeded with gels, we flip them and place them in an incubator to warm up and induce thermal sol-gel transition of the material for 15 minutes. Once complete, we removed the plates, added 100 µL of media into each well, placed 100 µL of PBS in surrounding, emptied wells as an evaporation buffer, then returned the plates to the incubator until performing the imaging study.

## 1.3 Cell Media

We used two different media, mouse splitting media (MSM) and the Complete Feeding Media (CFM), to culture PDAC spheroids as previously published[1]. MSM was made by combining then filtering 485 mL of advanced DMEM/F12, 5 mL of Pen/Strep, 5 mL of GlutaMax, and 5 mL of HEPES. MSM served as a base for CMF, which we made in 200 mL batches. We used 173 mL of MSM and supplemented the media with 20 µL each of A-83-01, mEFG, hFGF-10, and hGastrin I. Further supplementation included 200 µL of mNoggin, 500 µL of N-acetylcysteine, 2 mL of nicotinamide, 20 mL of R-Spo1 conditioned media, and 4 mL of B27.

## 1.4 Cell-laden Hydrogel Solutions

To prepare cell-laden hydrogel solutions, we first removed media using an aspirating pipette from wells containing previously cultured spheroids. The number of wells was pre-determined by how many wells must be seeded for the experiment. We then washed each well with 500 µL of Dulbecco's phosphate buffered saline (PBS) by running it down the side of the well to not disturb the hydrogel dome. We let PBS sit for up to 2 minutes before aspirating. Then we added 500 µL of Gentle Cell Dissociate Buffer to each well and scraped the wells using a 1000 µL pipette tip.

We collected all dissociated hydrogel and cell material from each well in a 15 mL conical tube. The tube was placed on a rocker in the 4 ºC cold room for 1 hour. Then, we mixed the solution using a p1000 pipette by manually resuspending the cell, hydrogel, and dissociation buffer solution 20 times. We spun down this solution at 300 rcf for 5 minutes. Once centrifuged, we aspirated the supernatant and added 1 mL of fragmenting solution (500 µL TrypLE, 500 µL PBS, 1 µL DNAse I) and gently agitated the solution for 5 minutes at room temperature. We quenched the trypsin reaction by adding three times the volume in the conical tube (3 mL) of cell media, mixed the solution, and then centrifuged the solution for another 5 minutes at 300 rcf. We then aspirated the solution above the cell pellet and resuspended it with 500 µL of cell media before straining the resuspended cells through a sterile 70 µm filter. We washed the initial tube with 500 µL media and strained it before counting the cells using a BioRad TC10 automated cell counter. We calculated the number of cells needed for the experiment and removed the appropriate aliquot from the cell solution before centrifuging for 5 minutes at 300 rcf. We aspirated the media and resuspended the cell pellet in the calculated amount of thawed hydrogel for direct plating or further use in nanoparticle gels.

## 1.5 Nanoparticle Hydrogel Solutions

When mixing solutions with nanoparticles, we first calculated the volume of nanoparticles needed from the stock solution to create a desired gel concentration at a specific volume. The volume is typically determined by the number of wells (10 µL of gel solution per well). For example, if our stock solution is 54 mg/mL, preparing 150 µL of gel (12 wells plus some extra) requires 2.78 µL of stock to obtain a final concentration of 1 mg/mL. We place the stock solution aliquot in a 1.5mL centrifuge tube and spin in a benchtop centrifuge for up to 5 minutes until the nanoparticles are separated from the initial solution. Then, we will remove the excess water using a micropipette, add the cell-laden hydrogel solution to the nanoparticles, and manually resuspend the particles. Extra care is taken not to introduce bubbles into the gel and to distribute the cells and nanoparticles evenly in the solution before plating in 10 µL increments in the middle of the desired well.

## 2  Dataset Generation

## 2.1 Overview

The spheroid samples were analyzed using 1) optical imaging and 2) human expert evaluation. To explore utility of SAAVY across microscopy platforms, imaging was performed on two different microscope systems: ECHO Revolve and Operetta CLS. Images were acquired on several different days during the growth process (Day 0 – Day 6). A pair of spheroid experts formally evaluated the images chosen for this study. A subset of samples were also analyzed using CellTiter Glo (CTG) assay. To further clarify the protocol, several terms should be defined.

First, two "Expert types" were involved in this work: spheroid expert and programming expert. Spheroid experts were chosen according to their training on 3D tissue culture protocols, viewing or imaging spheroids with microscopes, and familiarity with cystic PDAC spheroid systems in particular.

Second, four distinct image data sets were used: 1) programmer training data set, 2) SAAVY creation data set, 3) Echo 2D image set (mapping), and 4) Operetta 3D image set (longevity). All image data sets were of PDAC spheroids, and no fluorescent labels or stains were used. The SAAVY creation data set was divided into the SAAVY training data set and SAAVY validation data set. No images were re-used between these data sets. We intentionally isolated the spheroid experts from the training data set and kept the experts blinded to the metadata of the ECHO 2D image set.

Third, on day six, a subset of the samples imaged using the ECHO microscope were also analyzed using the CTG method.

## 2.2 2D and 3D Label-free Optical Imaging

Images of PDAC samples were taken for the study on an ECHO Revolve using an Olympus 4x lens on the day of seeding (D0), the day of chemotherapeutic treatment when applicable (D4), and on day 6 (D6) before developing the CTG assay. Spheroids were incubated the entire time and were only removed from the incubators for imaging or treatment.

For the longevity study, images were taken on an Operetta CLS with a 5x lens on all days of the assay (D0-D6). This instrument features an incubator chamber, and cells were maintained at 37 ºC and 5% $CO_2$ except for when they were transferred from the incubator to the instrument.

Additional details on the instrumentation and the protocols can be found in the experimental metadata section of this document.

## 2.3 Spheroid Expert Evaluation

The spheroid experts were instructed to take all spheroids in the image into account when assessing aggregate spheroid viability for the entire image. If the image appeared to have no spheroid structures, the experts were asked to note this fact. Experts were asked to time themselves during estimation to provide basis for efficiency comparison as well. They were not provided any information regarding growth day, perturbation condition, or matrix composition.

Images analyzed in the main text were randomly selected for inclusion. Images were renamed with a randomly generated image number index to blind the images before providing images for expert analysis. We further discuss expert variability and establishment of ground truth in subsequent sections.

## 2.4 CellTiter Glo Analysis

CellTiter Glo (CTG) is a luminescence-based viability assay commercialized by Promega. We use the procedure provided by Promega to conduct our analysis. Briefly, CTG reagent is removed from storage freezers to thaw overnight at 4 ºC before use in the assay. Well plates are removed from the incubator and the reagent is removed from the fridge to equilibrate to room temperature 30 minutes before development. Equal volume of CTG reagent to cell media is added into each well for development. The well plates are placed on a shaker for 5 minutes to physically disrupt the gel matrix, fragment the spheroids, and lyse the cells before a 20 minute wait step. During this time, the reagent reacts with lysed cellular material and the bioluminescent signal stabilized for measurement. The plates are then read on a BioTek plate reader, and the results are exported in Excel format for normalization.

The normalization process includes subtracting the average material background from control wells, which are typically hydrogel matrix and cell media from the experiment wells. Cell control wells are used as the positive control for completely alive cells. The average luminescent signal of cell control wells is used as the divisor for all experiment wells. All final reported CTG values are normalized in this manner.

## 3 SAAVY Development

### 3.1 Visual indicators of culture health

Fluorescent labels and stains are typically used to differentiate live from dead cells and organoids, particularly in AI-based systems. However, experienced users can often determine viability without these aids. This type of label-free analysis opens to the door for longitudinal studies. However, human experts can be subject to bias in their assessment. Therefore, the first step in developing a label-free image system is identifying the cues that experts are relying on and developing a system that removes the human subjectivity.

We surveyed a group of experienced 3D tissue culturists to determine the visual metrics that they use in assessing 3D spheroid or organoid culture health. The survey collected information on the culturist's years of experience and the specific disease models. As can be seen in Figure S1, the majority of those surveyed have worked with 3D cultures for greater than 3 years (62.5% combined). Other than pancreatic cancer 3D disease models, colon (colorectal), breast, and liver cancers were represented in culturist expertise.

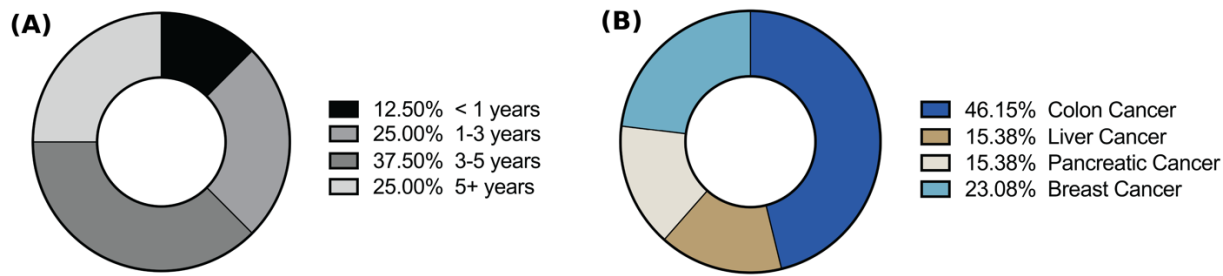

Figure S7: Pie chart summaries of the years of experience of the surveyed tissue culturists and the different disease types that the experts have analyzed.

The survey also asked the following open-ended question to determine the metrics that they assessed: "Think about healthy and unhealthy 3D tissue cultures you have observed in the past without cellular dyes. What visual metrics would you use to describe the culture health?" A summary of their responses, with percentage of surveyed experts who noted the metric, is provided in Table S1. The majority of experts noted two of the main metrics that we trained SAAVY to recognize: size and shape. They noted a relationship between the health of round (cystic) cultures and the "popped appearance" of dead cultures. The experts who noted transparency as a metric all indicated that darker cultures (opaque, less transparent) correspond to lower health. This bioindicator served as the cornerstone of SAAVY's predictive model.

Table S2: Summarized results of the metrics that surveyed experts provided in response to an open-ended question regarding visual indicators of 3D tissue culture health.

| Metric | % of respondents |
| --- | --- |
| Transparency | 75% |
| Morphology – shape | 63% |
| Media color | 38% |
| Morphology – size | 75% |
| Culture consistency | 13% |
| Density | 25% |

## 3.2 SAAVY Overview

An overview of the key elements of SAAVY is shown in Figure S2. Notably, all programming is performed in Python. PyTorch was chosen as our backend framework because of its speed and simplicity. The conda environment file is available within GitHub; however, instructions for installation differ slightly based upon the availability of a CUDA capable GPU. SAAVY was trained and tested on an Nvidia RTX 3080 with 12 GB of VRAM, Intel Core i9 10850K CPU. The images are rescaled during inference to fit into most GPUs with less than 4GB of VRAM.

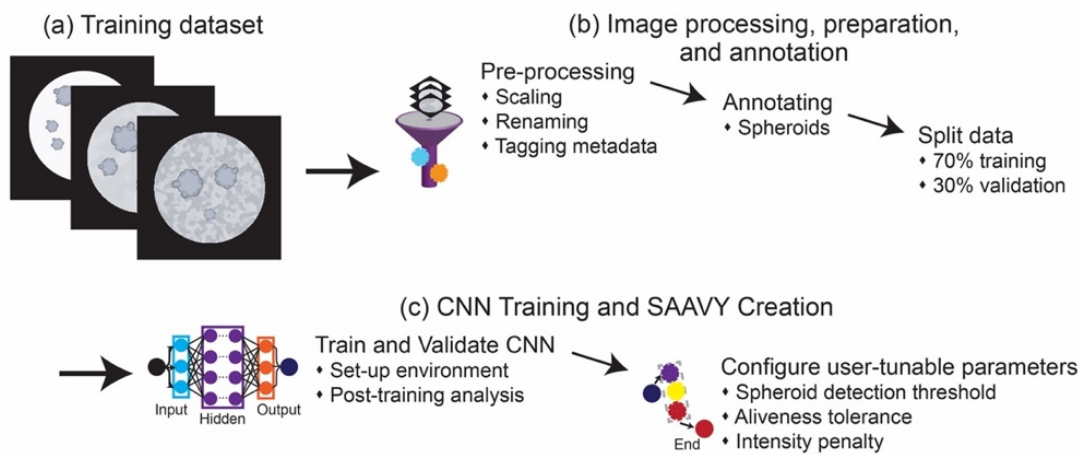

*Figure S8: Overview of SAAVY training process. a) 30 images were randomly selected for use as the training data set, b) the image processing step included rescaling the images to workable sizes and annotating 24 (70%) of the randomly selected training images and c) SAMY validation on the remaining 9 (30%) images of the data set and some fine-tuning of parameters for best performance.*

## 3.3 SAAVY Design

SAAVY is composed of two high level steps: spheroid segmentation and individual spheroid analysis. The segmentation step is a Mask R-CNN PyTorch model, which adds the ability to mask segmented instances for individual analysis of spheroids. Mask-RCNN is highly performant in instance segmentation and is based off Faster R-CNN[2] which predicts bounding boxes, and class scores for potential objects in an image. We elected to start from the MS COCO pretrained dataset to take advantage of transfer learning and refined the model from there. The box predictor and mask predictor were both replaced and trained on our dataset. We ran our analysis on each of the masked spheroids as outlined in Figure S2.

## 3.4 Training Dataset Creation

The training dataset was comprised of 24 images balanced between day (Day 4 and Day 6), status (alive and dead), and background type (clear and noisy), as noted in Table S2. Per the protocol, the data was resized to 1290 x 1210 (px). Then, the programming expert annotated each spheroid inside the image using VIA Image Annotator (2.0.11, Oxford). A class label was manually assigned to each spheroid where the "cell" class label is used for all detections. The saved annotations were exported in the JSON file format.

*Table S3: The table notes the distribution of images included in the training dataset. Of the 24 included, they were balanced by day (12 for each) and background type subset (6 for each day's background subset). We further break down the inclusion of sample viabilities (based on rough groupings) to note that we included representative stages of viability within the training for identifying spheroids.*

| Day 4 | | | Day 6 | | |
|---|---|---|---|---|---|
| Total | | 12 | Total | | 12 |
| Clear | | 6 | Clear | | 6 |
| | Dead | 0 | | Dead | 0 |
| | Mostly Dead | 0 | | Mostly Dead | 2 |
| | Mostly Alive | 3 | | Mostly Alive | 1 |
| | Alive | 3 | | Alive | 3 |
| Noisy | | 6 | Noisy | | 6 |
| | Dead | 1 | | Dead | 3 |
| | Mostly Dead | 0 | | Mostly Dead | 2 |
| | Mostly Alive | 3 | | Mostly Alive | 0 |
| | Alive | 2 | | Alive | 1 |

### 3.5 Model Training

Using supervised learning, SAAVY was trained using the annotated images for 20 epochs saving checkpoints and Tensorboard updates at the end of each epoch. The images should be scaled accordingly to fit into VRAM. Images with dimensions of 1200x900 fit into most 4 GB GPUs. Learning rate and similar training parameters were based on those used by Matterport's M-RCNN examples. Based on the validation loss and epoch loss, we used the checkpoint at epoch 15 when improvement stops (Figure S3).

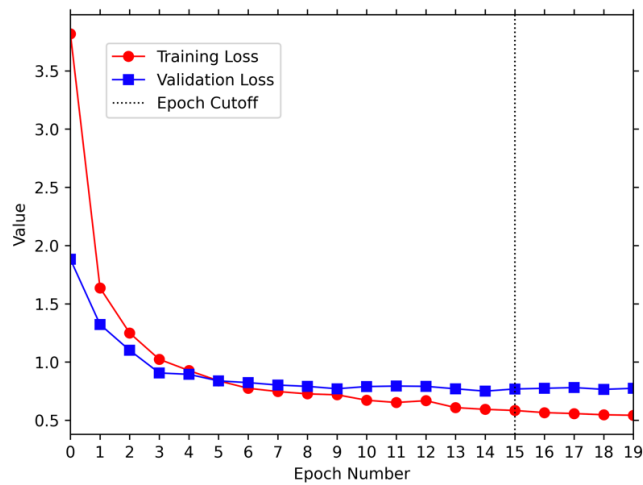

*Figure S9: Loss plot denoting the stabilization of loss around the same asymptote for both training and validation loss with the decided-upon epoch cutoff noted at 15 with a dotted line.*

## 3.6 Prediction and Analysis

As part of the algorithm, we developed an equation that compares the average intensity of the masked regions of interest against the overall background intensity of an image. Further, we included a configurable distance penalty that relates the pixel of interest to the overall background intensity and a minimum tolerance that serves as the minimum boundary condition on the intensity for appropriate viability assessment. Both values were determined by expert analysis and comparison during the initial setup. A lower distance penalty results in a smaller viability decrease per unit of intensity decrease, and the minimum tolerance allows for spheroids without complete transmittance to be classified as 100% alive. To determine the minimum tolerance, we noted the intensity of a 100% alive spheroid, as determined by expert analysis, and compared it to the background intensity of an image without nanoparticles.

The trained weights checkpoint was loaded using Pytorch, and we use that to inference each image. The output is a list of tagged spheroids and vertices of their masks. This information is passed to the next step, which creates a mask using the vertices and inverts it to calculate the mean background intensity. Then, it iterates through each set of spheroid masks to calculate the mean spheroid intensity. The eccentricity of each spheroid and its size are also determined at this step. The final viability of an individual spheroid is calculated by comparing the ratio of the spheroid intensity against the background intensity, as noted in Equation S1.

Equation S1 | $Viability = 1 - \frac{clamp(Background - Minimum\ Tolerance - Mean\ Value)}{Distance\ Penalty}$

Based on this expression, a healthy spheroid is mostly translucent whereas a dead spheroid is entirely opaque. This process was completed for each spheroid. The viability of the entire well is determined by averaging individual spheroid viability across the entire image. Finally, the vertices are drawn and connected to the original images with individual spheroid information.

## 3.7 SAAVY to Expert Identification Accuracy

To assess the accuracy of spheroid segmentation completed by SAAVY, we compared the expert count of spheroids to the total count per image output by SAAVY in twenty-five randomly selected images (Figure S4). All images in the mapping and longevity datasets were combined for the selection and the final images included ten from mapping and fifteen from the longevity set.

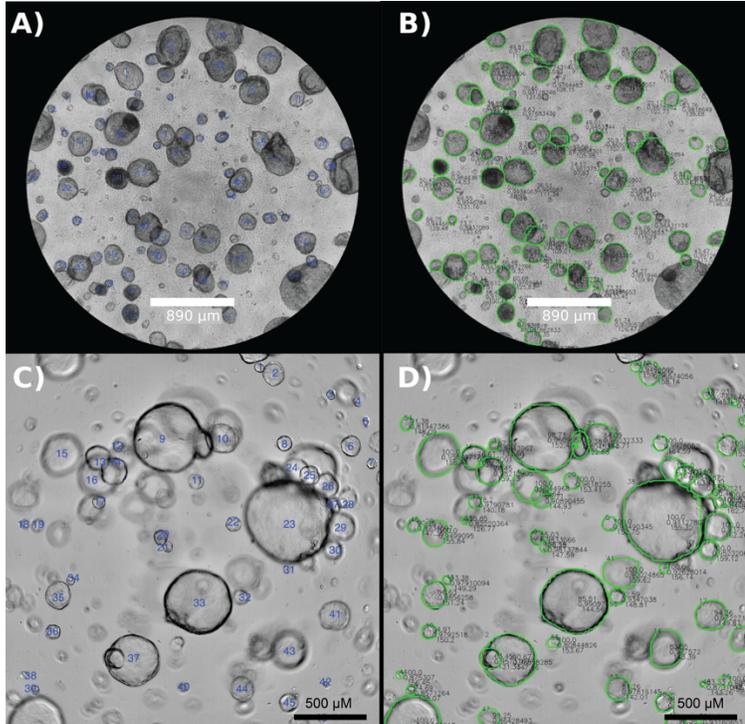

*Figure S10: Figure of spheroid expert and SAAVY counting comparison on an image taken using either the (A)/(B) Echo system or the (C)/(D) Operetta system. A) Image 409 from the Mappings dataset with expert labelling (blue numbers over spheroid). B) SAAVY segmentation (green outline) of Image 409. C) Image d3-320um_g08 from the planar longevity dataset with expert labelling (blue numbers over within spheroid). D) SAAVY segmentation (green outline) of Image d3-320um_g08.*

SAAVY and the expert did not predict differently when evaluating the average value of identified spheroids (p-value > 0.05). We further characterized the intersection over union (IoU, Equation S2) for each image to quantify the predicted overlap between SAAVY and expert.

$$\text{Equation S2} \mid IoU = \frac{True\ Positives\ (TP)}{True\ Positives\ (TP) + False\ Negatives\ (FN) + False\ Positives\ (FP)}$$

IoU falls between 0 and 1 (inclusive) where values close to 0 reflect model inaccuracy in the predicted label and a value of 1 reflects perfect model accuracy as compared to ground truth. We use the following definitions for IoU calculation: 1) TP: spheroids that SAAVY and Expert both identified, 2) FN: spheroids that SAAVY missed but expert identified, and 3) FP: spheroids that SAAVY predicts but expert does not. We calculated each image IoU in the random selection of images and report the average for each image grouping. For the 10 images from the mapping dataset (taken on the ECHO revolve) the average IoU is 0.622 and the average of the 15 images from the longevity study (taken on the Operetta CLS) is 0.653. Both IoU values are above 0.5 indicating alignment. Additionally, the results are consistent across imaging systems, indicating that SAAVY is compatible with both systems.

# 4  Analysis

## 4.1 Data Distribution and Expert Harmonization

In Figure S5, we display the data distributions for all spheroids estimated at ≥ 0 in 100 bins to provide the count of estimated viability in 1% increments. The experts showed patterns in how they assigned values. Specifically, they assigned viabilities in roughly 5% (for Expert 1) and 10% (for Expert 2) ranges. SAAVY clearly displays a distribution of viabilities across the entire range with minor gaps in increments that are more likely attributed to not having an image fall within the specified range.

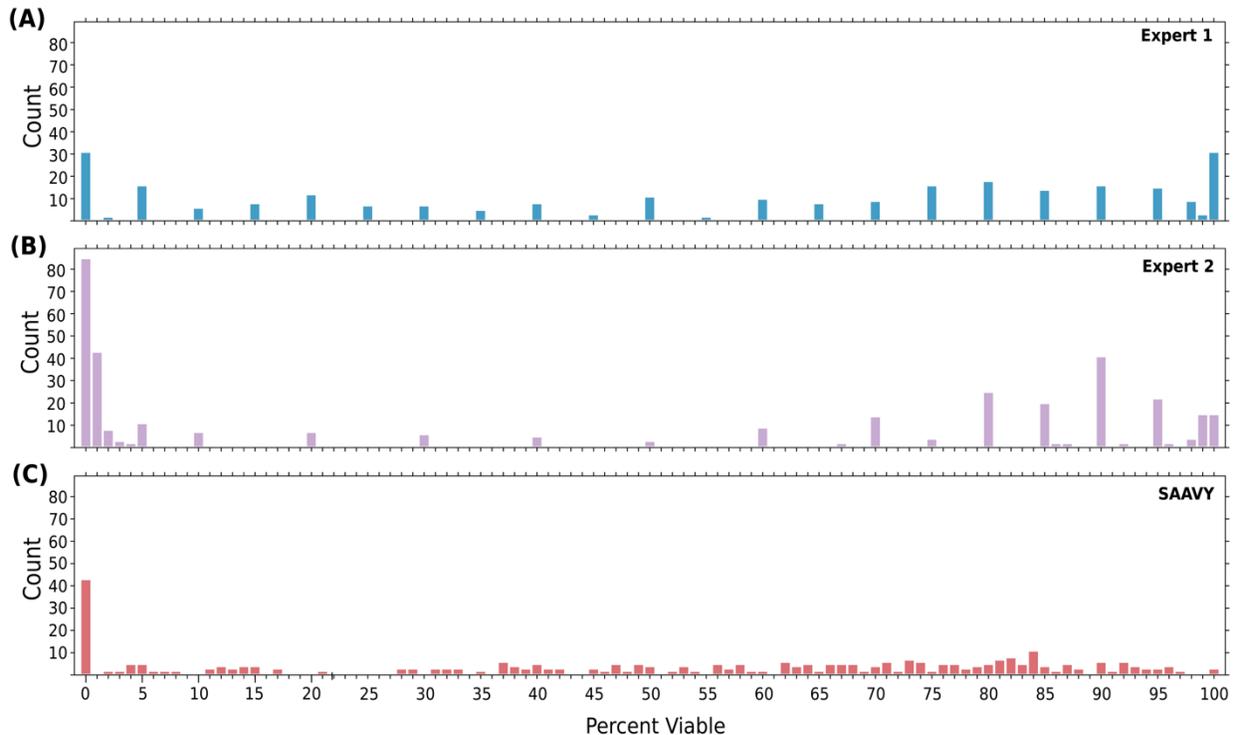

*Figure S11: Individual percent binning of a) Expert 1, b) Expert 2, and c) SAAVY analysis for the same set of images. Expert 1 shows ~5% gaps in estimation, Expert 2 displays ~10% gaps in estimation when far from the extreme edges of the distribution, and SAAVY displays a nearly even distribution without significant gaps in percent viability.*

This difference in assessment bin sizes and nomenclature will influence the accuracy of the subsequent statistical analysis. Therefore, we applied a commonly used data harmonization method to address the differences between expert estimation (Figure S6). First, we check the expert group distributions at the 5% bin increment level, the smallest of the two expert estimation gaps. With this even bin size, gaps were still present in Expert 2's distribution (Figure S6-A). This finding suggested that 10% viability increments for bins would be most appropriate to group subsequent distributions and analyses going forward. Next, all "none detected" labels were converted to 0% viability. This reduces the impact of Expert 2's inconsistent labelling and balances out the largest grouping differences present in the data. As absent spheroids cannot be assigned a viability, we assigned "0%" to mean an image may include either dead or absent spheroids.

Lastly, we check the expert distributions against 10% bin increments to confirm that the expected grouping appropriately converges the expert distributions. We confirmed this harmonization through a Tukey post hoc test for statistical differences among all group comparisons (Expert 1, Expert 2, and SAAVY) in all background types. Table S3 holds the expert-to-expert comparison between all data and group subsets. Although the Day 4 clear sub-grouping of data is statistically different, the overall distribution between Expert 1 and 2 is not significantly different and allows us to consider the experts holistically the same.

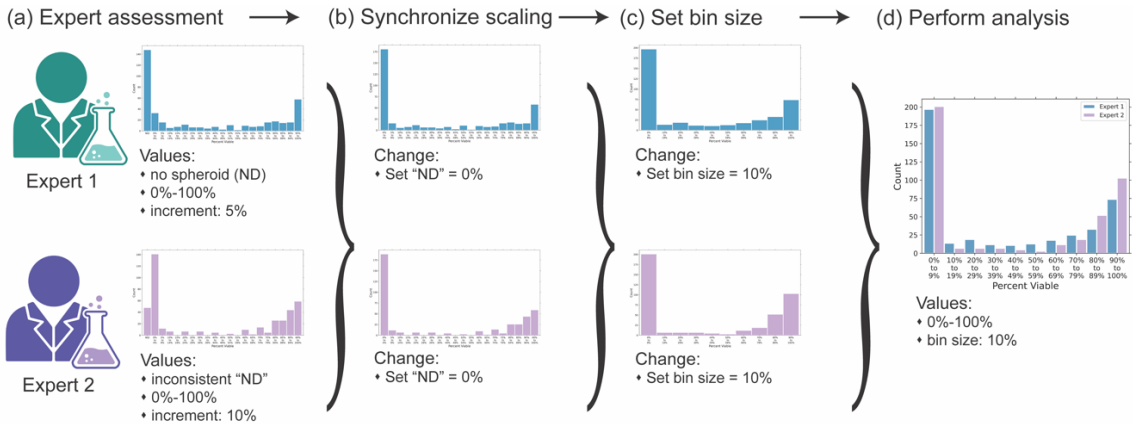

*Figure S12: Overview of the data harmonization method. The process is performed on all expert data sets. (a) A representative pair of expert assessments. Differences in assessment metrics that should be harmonized are noted. (b) The scaling is synchronized by merging the "none detected" spheroid group with the 0% viability estimate. Histogram bins are maintained at 5% increment levels. (c) The bin size is set to 10% increments. (d) Histogram of the harmonized expert data presented side-by-side.*

*Table S4: Calculated p-values from Tukey HSD post-hoc test for paired samples of expert-to-expert group differences. Red color is closer to 1 (not significantly different) and blue is closer to 0 (significantly different). All group subsets for day and background were analyzed and presented with the final row showing the overall data significance. All data and data subsets, except for Day 4 Clear, are not statistically different.*

|           | Expert 1 v. Expert 2 p-value |
|-----------|------------------------------|
| D0 Clear  | 0.999                        |
| D0 Noisy | 0.999 |
| D4 Clear | 0.0004 |
| D4 Noisy | 0.216 |
| D6 Clear | 0.520 |
| D6 Noisy | 0.861 |
| All Data | 0.167 |

### 4.2 Establishing Ground Truth

AI segmentation algorithms depend on ground truth data for appropriate model training and once trained, should be compared to ground truth to evaluate the overall model performance. One strategy is to have experts serve as ground truth. As we showed above, both experts predict viabilities in statistically similar manners. However, SAAVY provides viability values with more significant figures (viability per pixel is 0.4% compared to experts 1 and 2 who estimated in integer increments). Therefore, we wanted to test the spheroid experts further on their objective shade determination.

To do this, we wrote a program that would randomly pick a shade of gray and asked the experts to predict the percent white of the image (i.e., a 0% white image is black, and a 100% white image is white). The code recorded and output the expert-estimated value and the true percent value of the shade. Using the same distribution visualization and JSD calculations detailed in the main text, we found that Expert 1 more accurately assessed the true intensity value in an image (Figure S7).

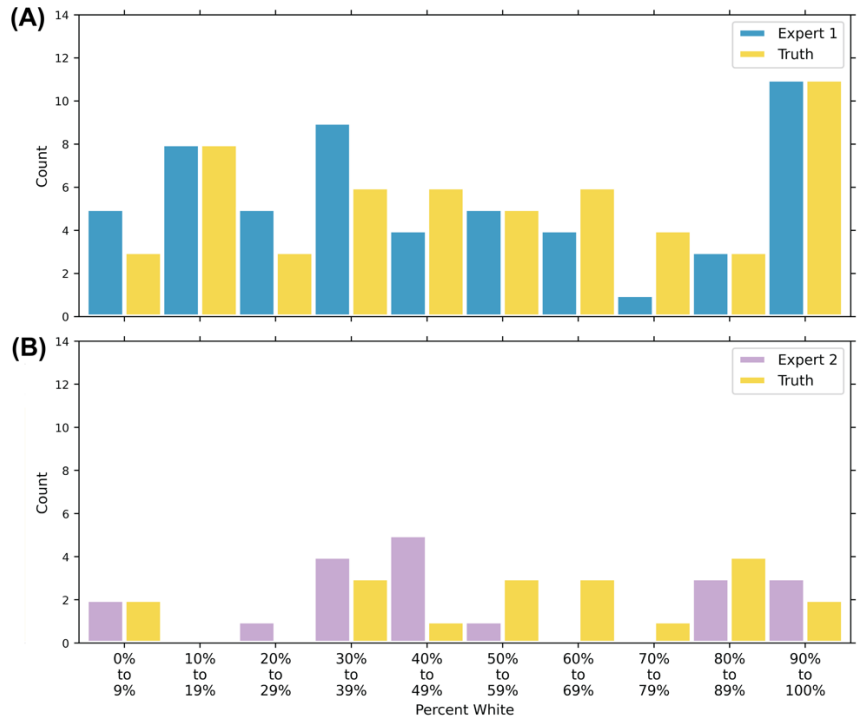

*Figure S13: Analysis of Expert 1 and Expert 2 greyscale assessment compared to ground truth values. a) Expert 1 and b) Expert 2.*

Expert 1 has a lower Jensen Shannon distance (0.140577) than Expert 2 (0.386908), suggesting that the distribution of Expert 1 is closer to that of the truth, although both are considered significantly different distributions than the respective truth. Expert 2 displayed gaps in prediction between 60% to 80%, thus missing a large portion of grey value differences, whereas Expert 1 estimated within the entire range of binned valued. Therefore, we decided to move forward with Expert 1 as our ground truth for the classification of SAAVY.

## 4.3 Relevant Equations for the Confusion Matrix

We use a confusion matrix to classify SAAVY analyzed images into true positive (TP), false positive (FP), true negative (TN), and false negative (FN) groupings. In TP images, SAAVY and spheroid expert pass the same classification. For example, in healthy spheroid images, both will assign a viability based on the implicit classification of live biologics in the image. FP images are those that SAAVY assigns a viability to that spheroid expert notes 0% viability (dead classification) or passes to a none detected category (for no identifiable spheroids in the image). TN images are those that SAAVY and expert assign a 0% viability or a "none detected" class. FN images are those where SAAVY assigns a 0% viability or a "none detected" class where expert assigns a viability (classifies as live).

Equations S3 and S4 were used to characterize the performance of SAAVY at identifying spheroid/no spheroid and live/dead spheroids throughout the analysis.

Equation S3 | $Accuracy = \frac{TP+TN}{TP+TN+FP+FN}$

Equation S4 | $F1\ score = 2 * \frac{precision*recall}{precision+recall}$

Equation S5 | $precision = \frac{TP}{TP+FP}$

Equation S6 | $recall = \frac{TP}{TP+FN}$

## 4.4 Distance Metrics

We compared Jensen Shannon Distance (JSD), Kullback Leibler Divergence (KL), Earth Mover's Distance (EMD), and Population Stability Index as distance metrics. JSD, KL, and EMD values depend strongly on the bin width chosen, as the data was discretized in a histogram before use. We calculated the ideal number of bins according to the expert harmonization section discussed above and that 10% increments bin our experts to statistically similar distributions. Due to the output of "none detected" in the SAAVY-output predictions coded as "-1", we used 11 total bins from -10 to 100 (inclusive).

We also included Krippendorff's alpha, a reliability metric, to evaluate systemic differences between methods of viability estimation for the experts and SAAVY comparisons. The results of these comparisons are included in Table S4.

*Table S5: Summary of all statistics run to compare various distance metrics against each other during exploratory analysis (columns Jensen Shannon – Population Stability Index). All calculations are completed as SAAVY compared to Expert and can be interpreted as Expert distribution being a certain distance from the SAAVY distribution. All metrics except for Kullback Leibler are also mathematically symmetric and can be interpreted as Expert distribution compared to SAAVY. Inf in an abbreviation for infinity, which is achieved when there are 0 data within a given bin of the data distribution. Values closer to 0 indicate distributions that more similar.*

|  |  |  | Jensen Shannon | Kullback Leibler | Earth Mover's | Population Stability Index |
|---|---|---|---|---|---|---|
| Day 0 | Clear | Expert 1 | 0.470 | 0.717 | 0.094 | 2.983 |
| | | Expert 2 | 0.711 | 2.261 | 0.086 | 2.693 |
| | Noisy | Expert 1 | 0.548 | 1.052 | 0.119 | 0.279 |
| | | Expert 2 | 0.406 | 0.515 | 0.073 | 1.034 |
| Day 4 | Clear | Expert 1 | 0.334 | inf | 0.029 | 1.469 |
| | | Expert 2 | 0.533 | 1.163 | 0.073 | 3.154 |
| | Noisy | Expert 1 | 0.266 | 0.312 | 0.021 | 1.035 |
| | | Expert 2 | 0.291 | 0.348 | 0.036 | 0.799 |
| | Clear | Expert 1 | 0.384 | inf | 0.045 | 2.724 |
| | | Expert 2 | 0.279 | 0.292 | 0.048 | 0.665 |
| Day 6 | Noisy | Expert 1 | 0.253 | 0.332 | 0.016 | 0.584 |
| | | Expert 2 | 0.220 | 0.163 | 0.011 | 0.828 |

## 4.5 Exploratory Comparison Analysis

Pearson's correlation was used to compare SAAVY to Experts and CTG and is plotted as heatmaps in Figure S8. CTG comparisons are only included where appropriate, as we only have matched samples on Day 6. For the Day 0 and Day 4 data sets, CTG was not included on the correlation matrices for clarity of the plot.

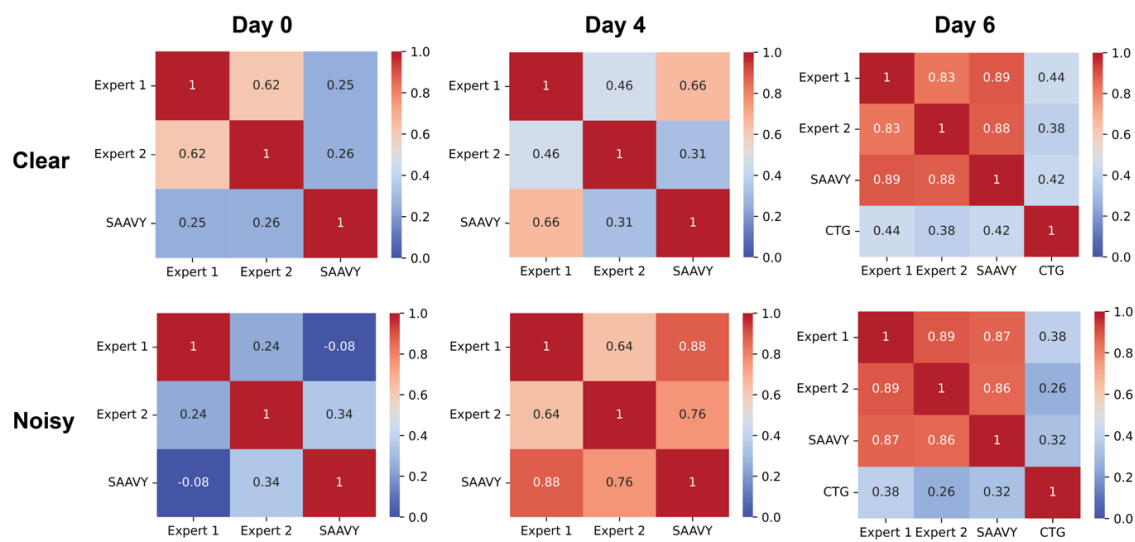

*Figure S14: Heatmap plots of Pearson's correlation between Expert 1, Expert 2, SAAVY and CTG (where appropriate). The plots are subset according to background type with clear backgrounds on the top row and noisy backgrounds on the bottom. Columns representing increases in day (from right to left). Red indicates improved correlation and blue indicates poor correlation.*

## 4.6 Exploratory Distribution Visualization

In Figure S9, we plot the distributions of the Expert to SAAVY comparisons from the main section of the manuscript. This reliability, however, improves on Day 4 and Day 6 as we note the convergence of the distributions between both experts and SAAVY.

The breakdown of CTG data as compared to SAAVY in combined, clear, and noisy background subsets is presented in Figure S10. In direct comparison, CTG values over 100% do not seem intuitive. However, it is important to state that these values are relative to the plate, and they make sense within the specific assay the results belong to. Note that CTG cannot produce values of 0 or below, and SAAVY only reports 0-100%.

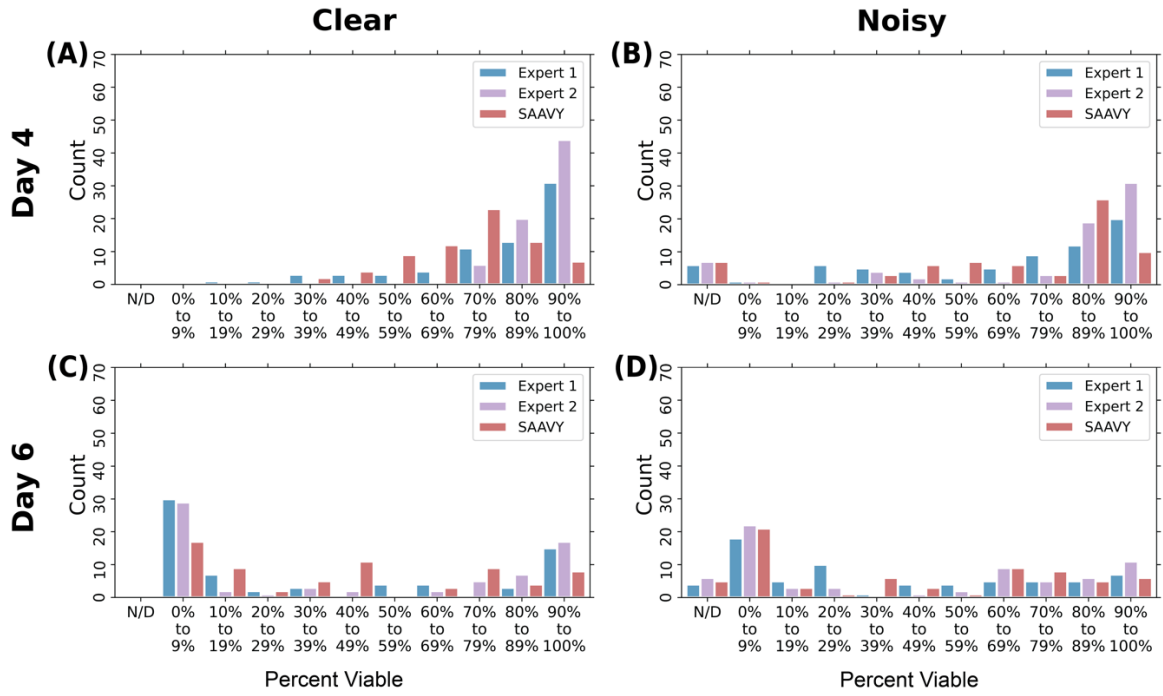

*Figure S15: Histogram distributions of a) Clear Backgrounds comparing Experts 1 and 2 and SAAVY on Day 4, b) Noisy Background viabilities predicted by Expert 1 and 2 and SAAVY on Day 4, c) Clear Backgrounds comparing Experts 1 and 2 and SAAVY on Day 6, d) Noisy Background viabilities predicted by Expert 1 and 2 and SAAVY on Day 6, e) Clear Backgrounds comparing Experts 1 and 2 and SAAVY on D6, f) Noisy Background viabilities predicted by Expert 1 and 2 and SAAVY on D6. N/D denotes "none detected".*

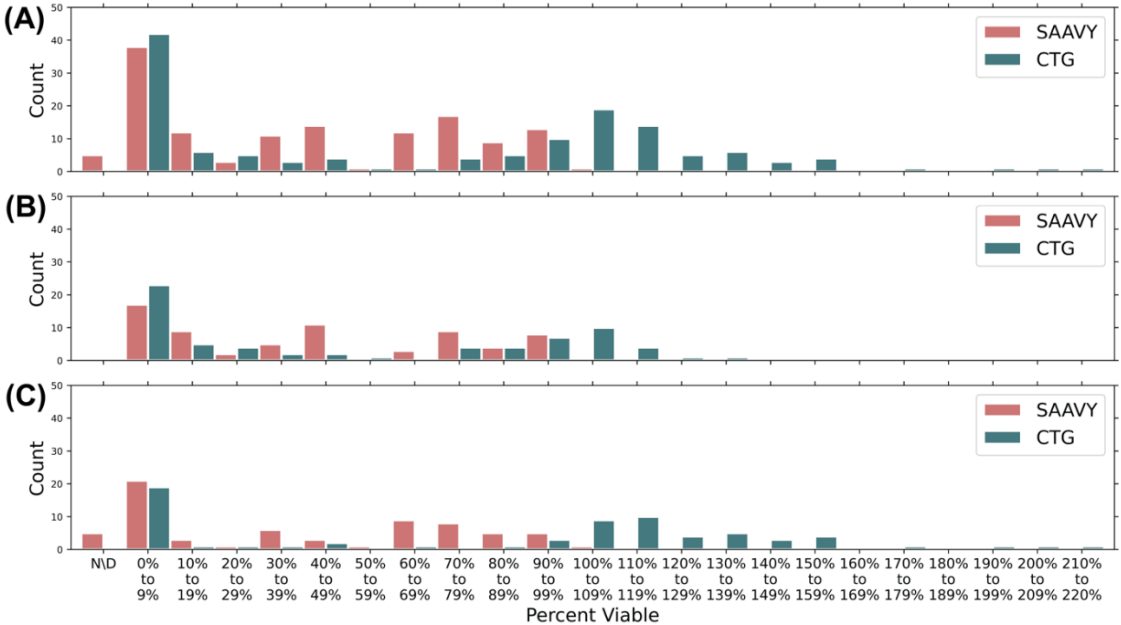

*Figure S16: Histogram plots comparing SAAVY and CTG performance across all matched samples in a) the combined dataset, b) the clear subset, and c) the noisy subset. Critically, we display the frequent error of CTG to report greater than 100% viability, and when used in samples that have noisy backgrounds when normalized (with included material standards to adjust for background noise, where appropriate).*

## 4.7 Individual Spheroid Analysis

The longevity study presented in the main text is the extracted whole-well information. We expand upon that analysis and study the results at the individual spheroid resolution level. In the SAAVY code, we can toggle on the export of csv files for each image that includes the viability and associated information on every single spheroid within the image. We perform this process for all images analyzed in the main text and summarize the results here.

We present the viability in a 2D histogram against the day of the image taken for all perturbation concentration groups in Figure S11. The histogram is flattened and represented by the color map, where lower counts of spheroids in each bin is blue and higher counts is red. The untreated group shows individual spheroid viability is relatively constant across days with a slight drop in day 6 likely due to exhausted media nutrients. In plots S11 c-f, the sharp viability decrease after day 4, due to the introduction of the GEM perturbation, is clear.

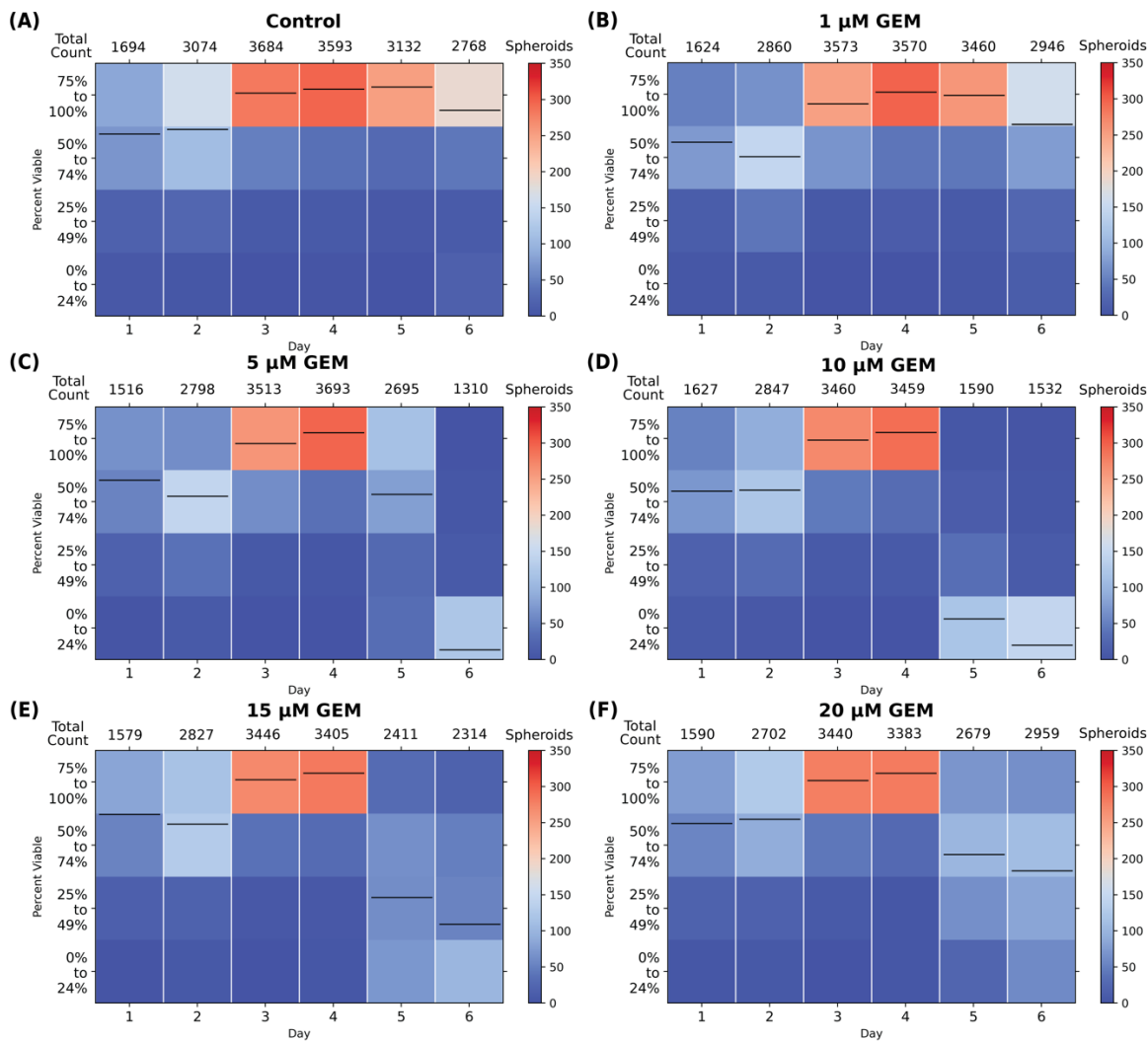

*Figure S11 - Histograms summarizing the number of spheroids within each viability bin (grouped in 25% increments) separated by day. The histograms are collapsed and presented in a color scale from 0 (blue) to 350 (red). Plots are separated by the concentration of added perturbation compound, gemcitabine (GEM). A) No added GEM. B) 1 µM GEM. C) 5 µM GEM. D) 10 µM GEM. E) 15 µM GEM. F) 20 µM GEM.*

We also track the area change over time in Figure S12. It is particularly interesting, as a shift in area distributions is not necessarily correlated to changes in viability. It appears that after day 4, spheroid growth is limited and even the introduction of the GEM treatment does not shift the larger area distribution. This suggests that area is unchanged with treatment. This is likely because the spheroids that are killed stay suspended in the gel matrix in a clumped together manner. Although the cells within the spheroid are no longer viable, the overall size is unchanged.

However, there is a difference in the distributions between the middle and higher concentrations in the spheroid response. Specifically, on day 4, all samples have a bimodal distribution of spheroid area, indicating that they are growing in the same manner as the control data. However, by day 6, the middle two concentrations (5mm and 10mm) have collapsed to a single size whereas the higher concentrations (15mm and 20mm) still maintain their bimodal shape.

While understanding the biology and biochemistry behind this behavior is outside of the scope of this work, it does demonstrate the potential utility of this approach in revealing mechanistic insights.

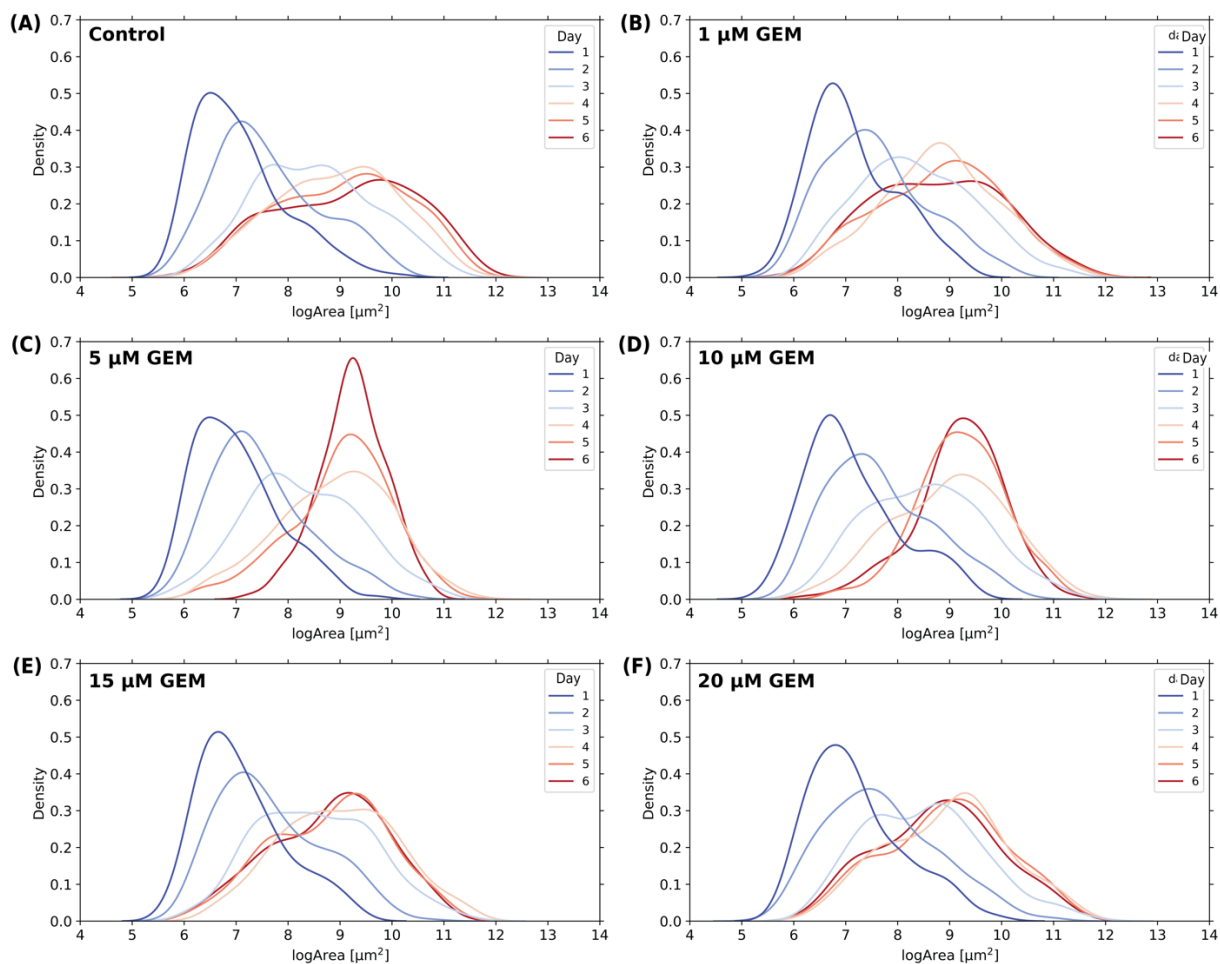

Figure S12 - Smoothed distribution (gaussian kernel density estimate) plots showing the change in area density over each day. Plots are separated by the concentration of added perturbation compound, gemcitabine (GEM). A) No added GEM. B) 1 µM GEM. C) 5 µM GEM. D) 10 µM GEM. E) 15 µM GEM. F) 20 µM GEM.

We investigate the change in these distributions further in the following set of figures S13-17. We plot and compare the viability before perturbing the system on day 4 to the same spheroid group viability on day 6. It is important to note that the individual points are index-matched spheroids from day 4 to day 6 and that, though the index matches, they are not necessarily the same spheroid. However, the visualization that we present agrees with the distribution shift observed in Figure S12. By comparing the same group against itself, we probe the dynamics of the system. All plots include a gaussian kernel density estimate (KDE) that we use a red coloring to note higher density of spheroids in the overall distribution. The black diagonal line in each plot is a line of no change. The region above the line is one of positive change, either of increased viability or area, and the region below the line is one of negative change.

The viability plots display the same trends observed in the population-level data. Specifically, the viability of the control sample is tightly clustered near 100%. In contrast, the viability of the GEM-treated sample is no longer tightly clustered but exhibits a broad distribution of values.

The area plots, however, point toward the interesting collapse of the bimodal distribution noted from Figure S12. The control group shows a bimodal distribution of both growth and decline, though the KDE coloring points toward higher density of spheroids that have grown over the course of the two days. Two perturbation groups are worth noting: the 1 µm and 5 µm. For the 1 µm group in Figure S13, we see a similar bimodal distribution in growth pattern of the spheroids even with a slight decrease in spheroid viability. In Figure S14, we see the overall collapse of the area distribution as compared to the control group. Most likely, the presence of a bimodal population and the change in distribution can be attributed to a variety of mechanisms, but performing a rigorous assessment is outside of the scope of this work.

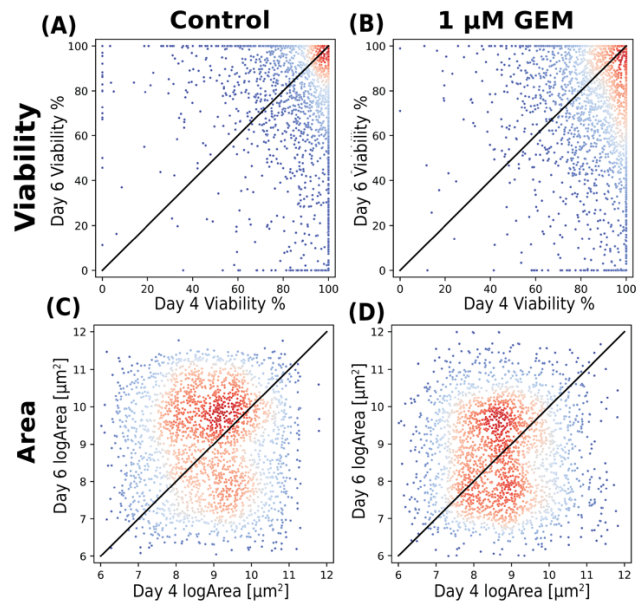

*Figure S13 - Scatter plots of individual spheroids from Day 6 plotted against Day 4 spheroids colored according to a gaussian kernel density estimation where blue is lower density and red is higher density. The plots either present the viability (A and B) or the log(Area) (C and D) for either the control group or the 1 µM gemcitabine (GEM) concentration perturbation group.*

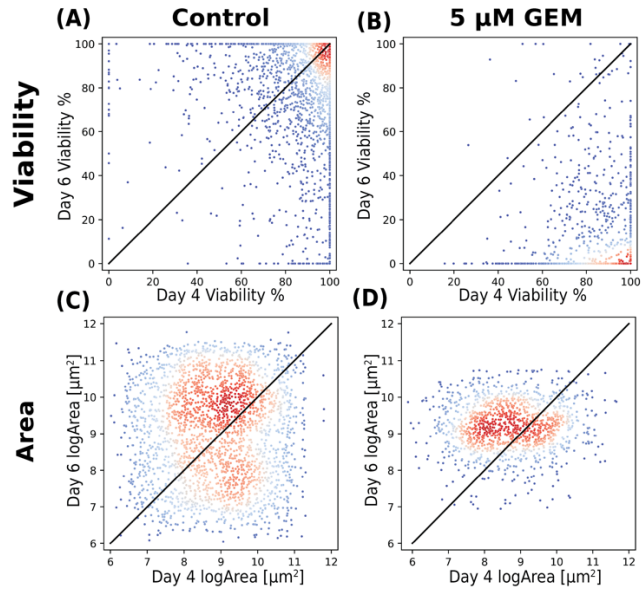

*Figure S14 - Scatter plots of individual spheroids from Day 6 plotted against Day 4 spheroids colored according to a gaussian kernel density estimation where blue is lower density and red is higher density. The plots either present the viability (A and B) or the log(Area) (C and D) for either the control group or the 5 µM gemcitabine (GEM) concentration perturbation group.*

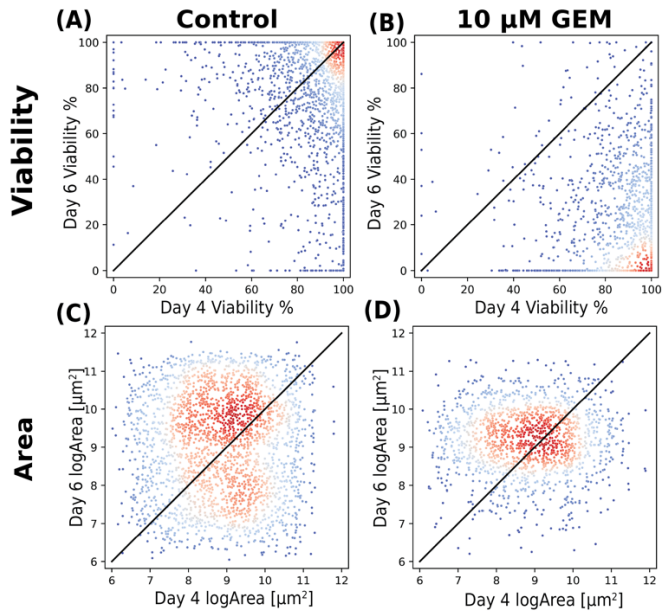

*Figure S15 - Scatter plots of individual spheroids from Day 6 plotted against Day 4 spheroids colored according to a gaussian kernel density estimation where blue is lower density and red is higher density. The plots either present the viability (A and B) or the log(Area) (C and D) for either the control group or the 10 µM gemcitabine (GEM) concentration perturbation group.*

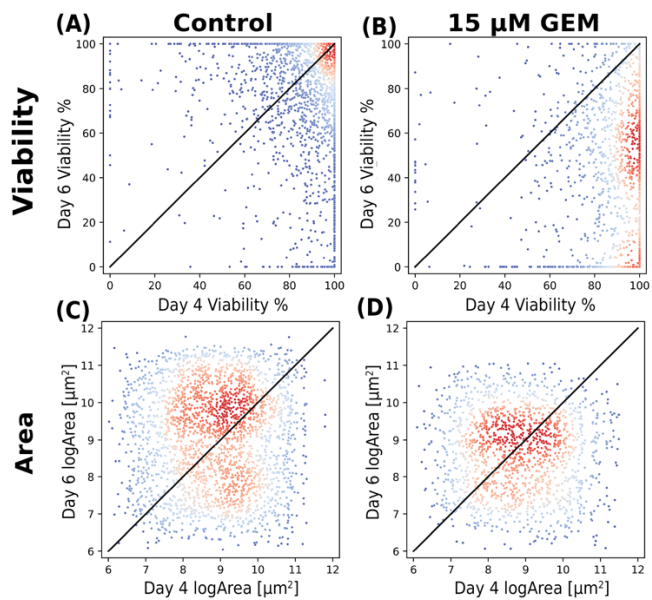

*Figure S16 - Scatter plots of individual spheroids from Day 6 plotted against Day 4 spheroids colored according to a gaussian kernel density estimation where blue is lower density and red is higher density. The plots either present the viability (A and B) or the log(Area) (C and D) for either the control group or the 15 µM gemcitabine (GEM) concentration perturbation group.*

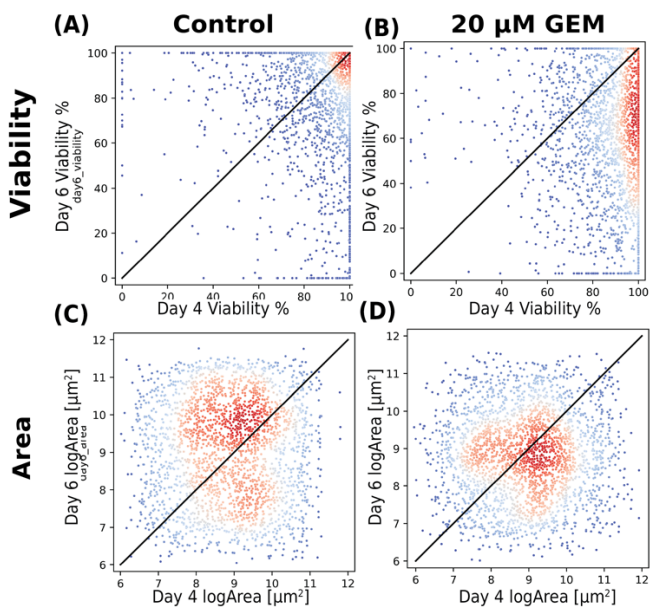

*Figure S17 - Scatter plots of individual spheroids from Day 6 plotted against Day 4 spheroids colored according to a gaussian kernel density estimation where blue is lower density and red is higher density. The plots either present the viability (A and B) or the log(Area) (C and D) for either the control group or the 20 µM gemcitabine (GEM) concentration perturbation group.*

## 4.8 Consistency in Multiple Planes

To test SAAVY's ability to predict viabilities of all spheroids, regardless of position within the gel and/or well, we analyzed the uniformity of the spheroid culture throughout the entire gel at the single-spheroid level. When running the longevity perturbation experiment, we took z-stack images. In the main text, only the mid-plane (320 µm) images were reported in the summary of results. Z-stack images were taken every 20 µm from 0 µm (the hydrogel/well-plate interface) to 700 µm (the very top of the gel where nothing was in focus). As part of this analysis, we used a subset of these planes that are spaced 40 µm apart (Figure S18).

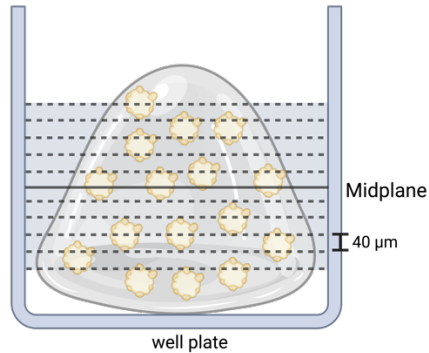

*Figure S18 - Figure depicting the planes that were selected for analysis in the planar reconstruction. Figure created with biorender.com.*

We classify the uniformity of SAAVY predictions over the viability and the area, presented in 2D histograms where the plane is the y-axis and the histogram is collapsed into a heatmap according to absolute count of spheroids within each bin. Every image exported for this analysis (2,376 in total) were analyzed according to each individual spheroid, with a single csv file summarizing the image exported for each. We combined the csv files into one using the cat *.csv >combined.csv terminal command on the resultant output folder. This information was cleaned up and metadata was added to the file to track perturbation concentration grouping.

Figures S19-24 show the uniformity of SAAVY viability estimates throughout the entire gel and shows the sensitivity of the algorithm to track changes over time. This is noted through the reduction in viability on Day 6, likely due to reduced nutrients in the cell media.

Figure S25-30 depict the changes in the area (reported on a logarithmic scale) over time within each plane. We see the area increase over time, as we would expect, plus the added resolution of planar differences between relative size of the identified spheroid. Although there are some differences in the distributions, the average areas calculated for each bin and depicted by the horizontal black lines within each plane group are similar for each grouping. This highlights the consistency of our algorithm regardless of plane.

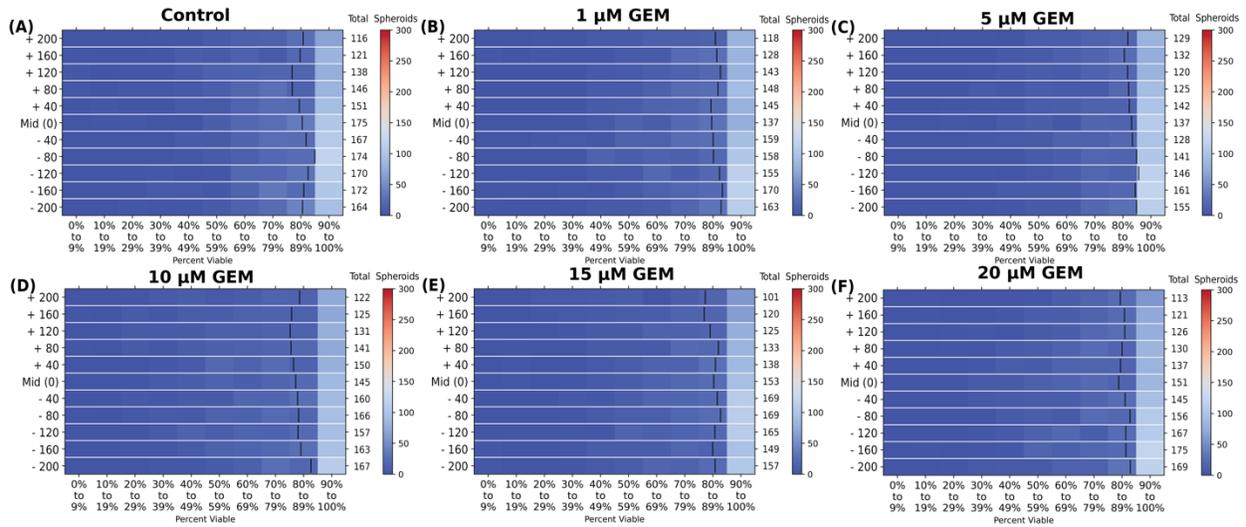

*Figure S19 – Day 1 2D histograms of spheroids separated by the focal plane from imaging (μm of distance from the bottom of the gel and well plate) against the percent viability (binned in 10% increments) of all individually identified spheroids in the data set. The total number of spheroids identified per each plane are noted in the "total" column between the plot and the color bar. The black bar within each plane grouping is the group's average viability. The color bar represents the total number of spheroids within each viability bin, where red colors indicate high counts (closer to 300) and blue indicate low counts (closer to 0) of spheroids within each of the viability bins. Data are segmented by day accordingly. Subplots are for the perturbation groups: A) control, B) 1 μM GEM, C) 5 μM GEM, D) 10 μM GEM, E) 15 μM GEM, F) 20 μM GEM.*

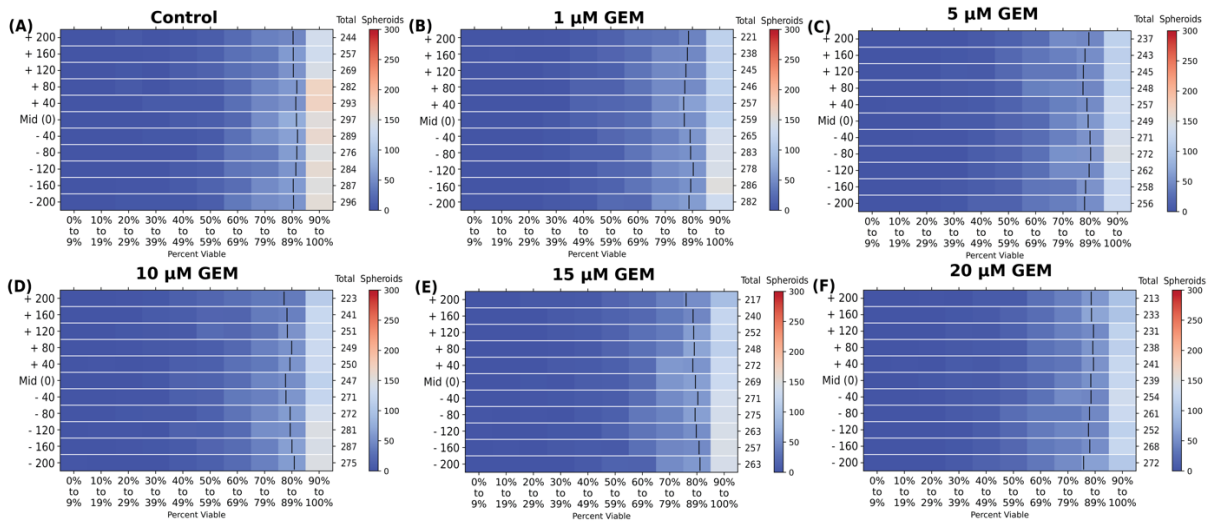

*Figure S20 - Day 2 2D histograms of spheroids separated by the focal plane against the viability bin.*

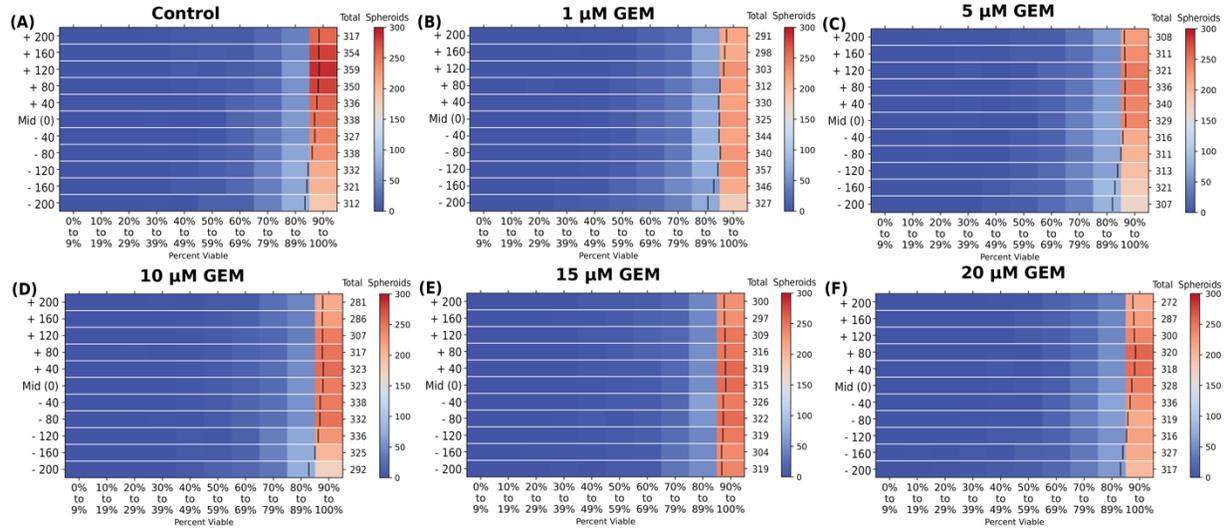

*Figure S21 - Day 3 2D histograms of spheroids separated by the focal plane against the viability bin.*

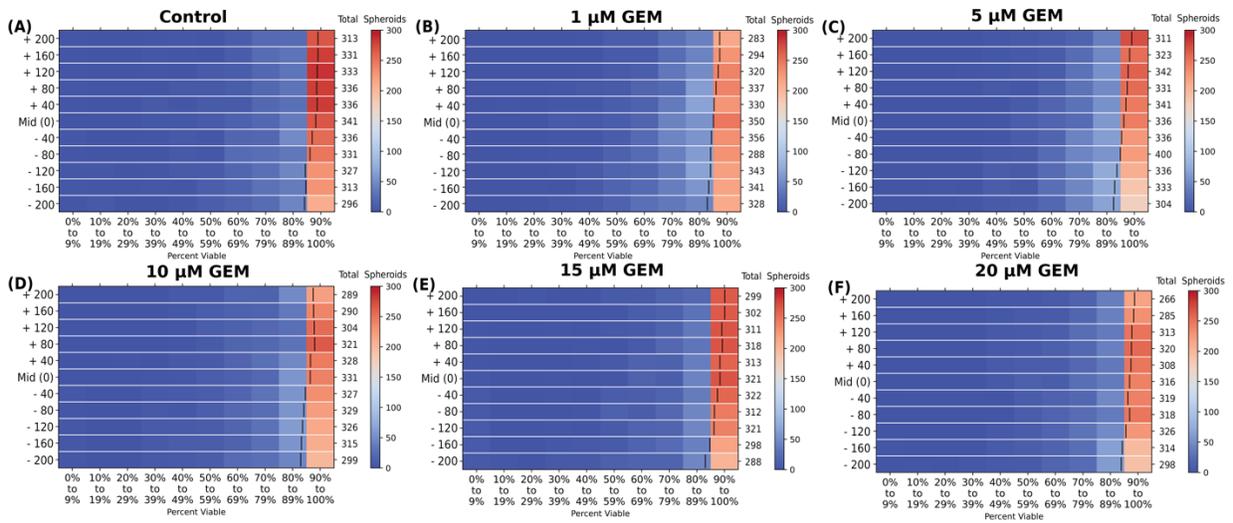

*Figure S22 - Day 4 2D histograms of spheroids separated by the focal plane against the viability bin.*

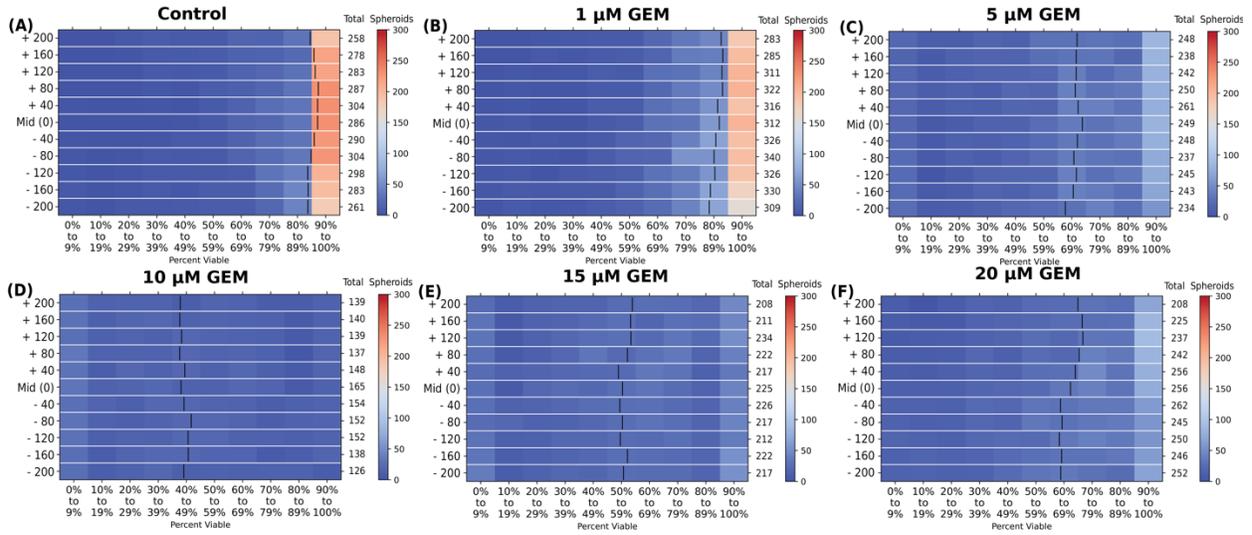

*Figure S23 - Day 5 2D histograms of spheroids separated by the focal plane against the viability bin.*

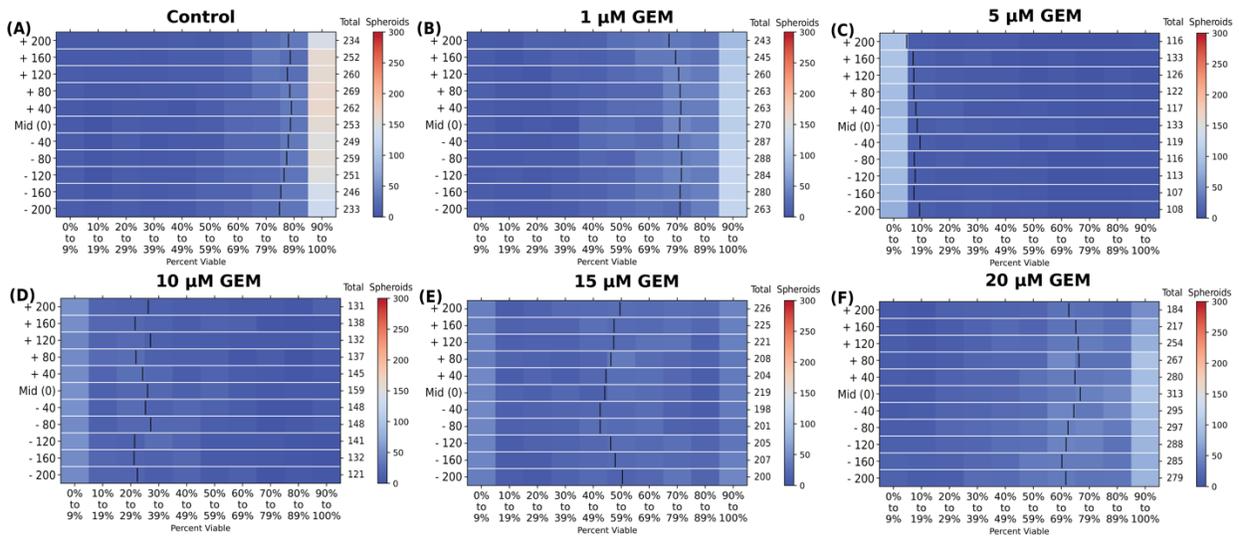

*Figure S24 - Day 6 2D histograms of spheroids separated by the focal plane against the viability bin.*

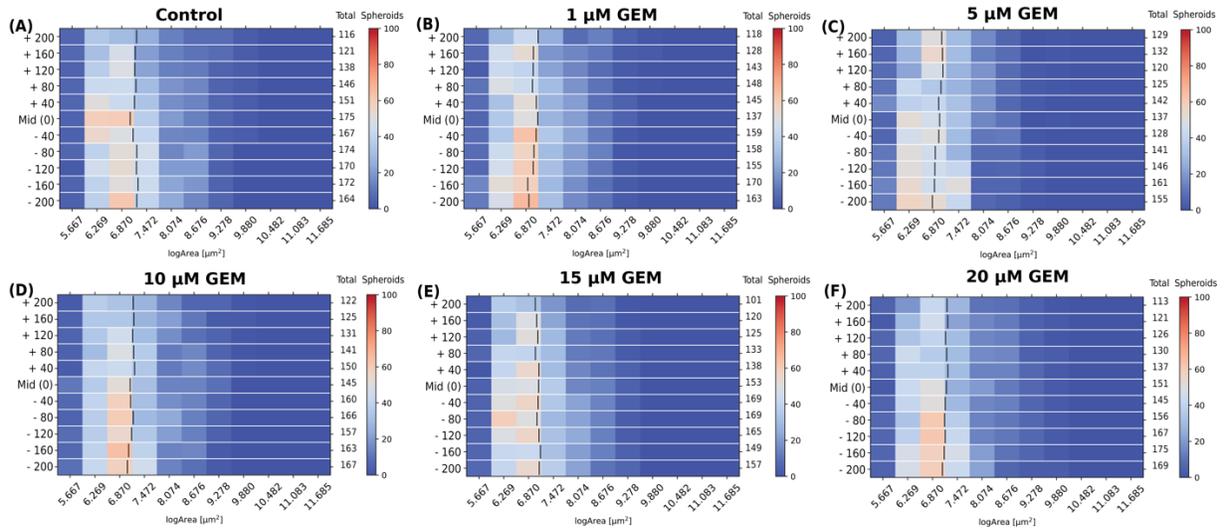

*Figure S25 – Day 1 2D histograms of spheroids separated by the focal plane from imaging (µm of distance from the bottom of the gel and well plate) against the log(Area) of all individually identified spheroids in the data set. The total number of spheroids identified per each plane are noted in the "total" column between the plot and the color bar. The black line within each plane group indicated the group's average area. The color bar represents the total number of spheroids within each viability bin, where red colors indicate high counts (closer to 300) and blue indicate low counts (closer to 0) of spheroids within each of the viability bins. Subplots are for the perturbation groups: A) control, B) 1 µM GEM, C) 5 µM GEM, D) 10 µM GEM, E) 15 µM GEM, F) 20 µM GEM.*

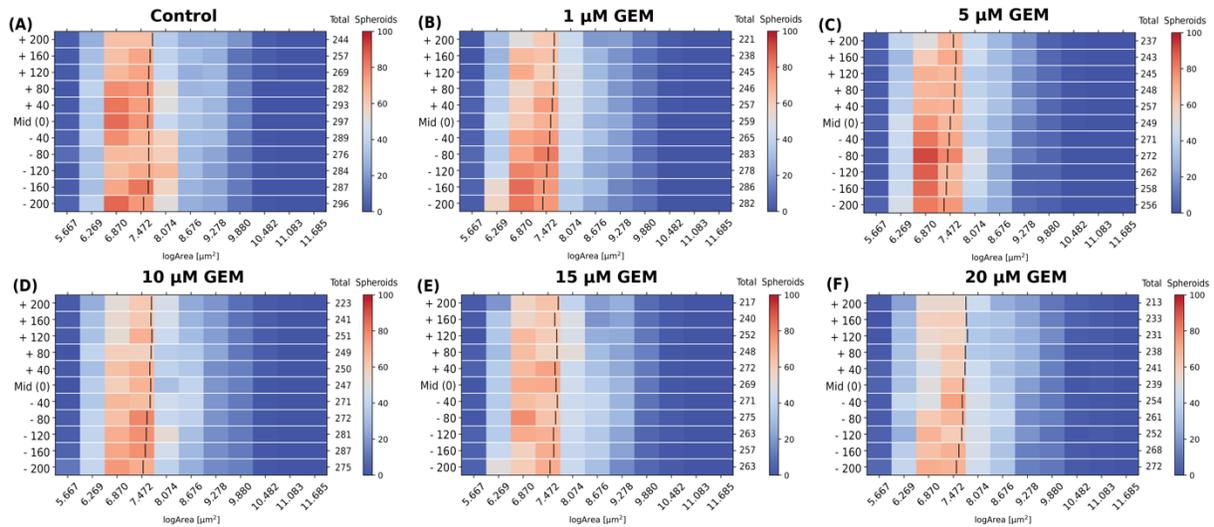

*Figure S26 - Day 2 2D histograms separated by the focal plane against the log(area) of all individually identified spheroids.*

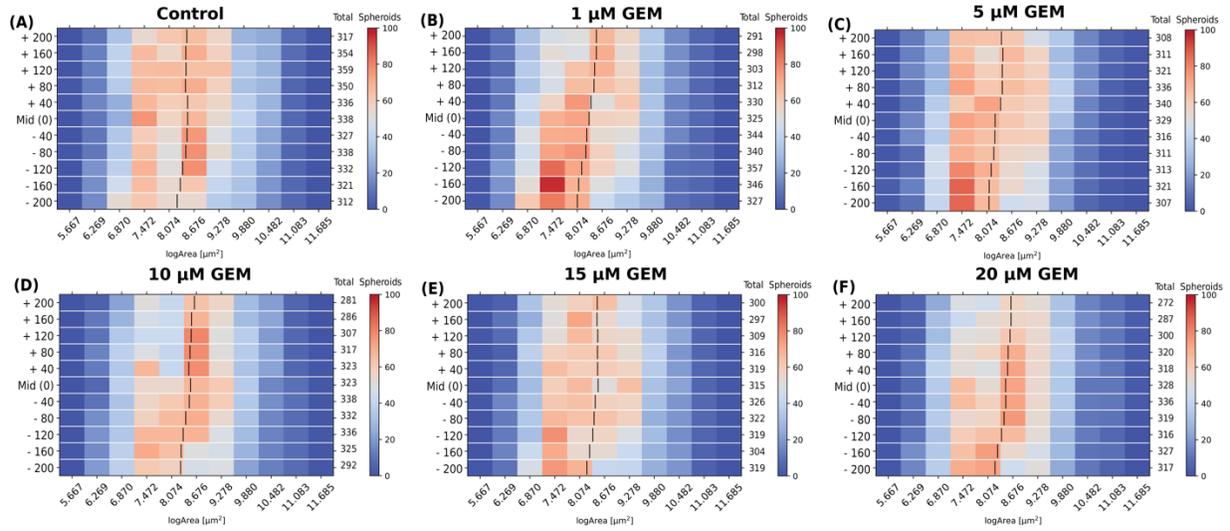

*Figure S27 - Day 3 2D histograms separated by the focal plane against the log(area) of all individually identified spheroids.*

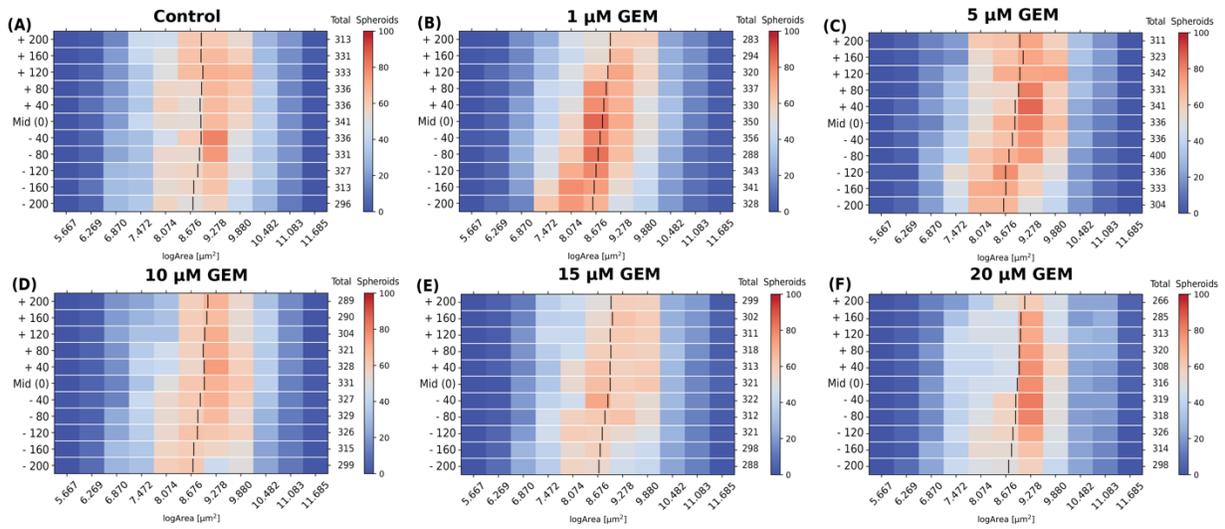

*Figure S28 - Day 4 2D histograms separated by the focal plane against the log(area) of all individually identified spheroids.*

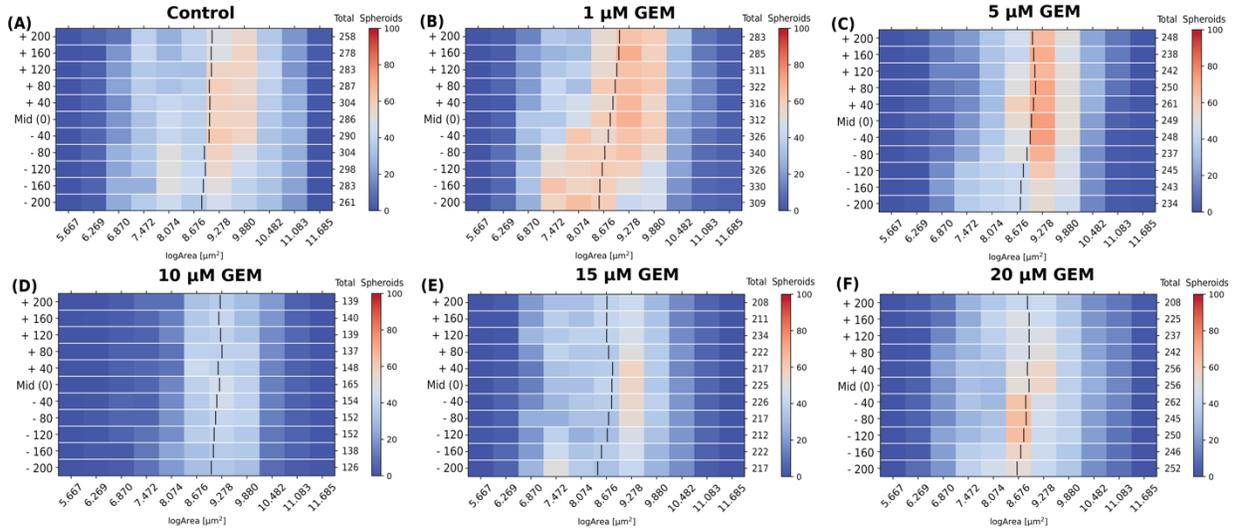

*Figure S29 - Day 5 2D histograms separated by the focal plane against the log(area) of all individually identified spheroids.*

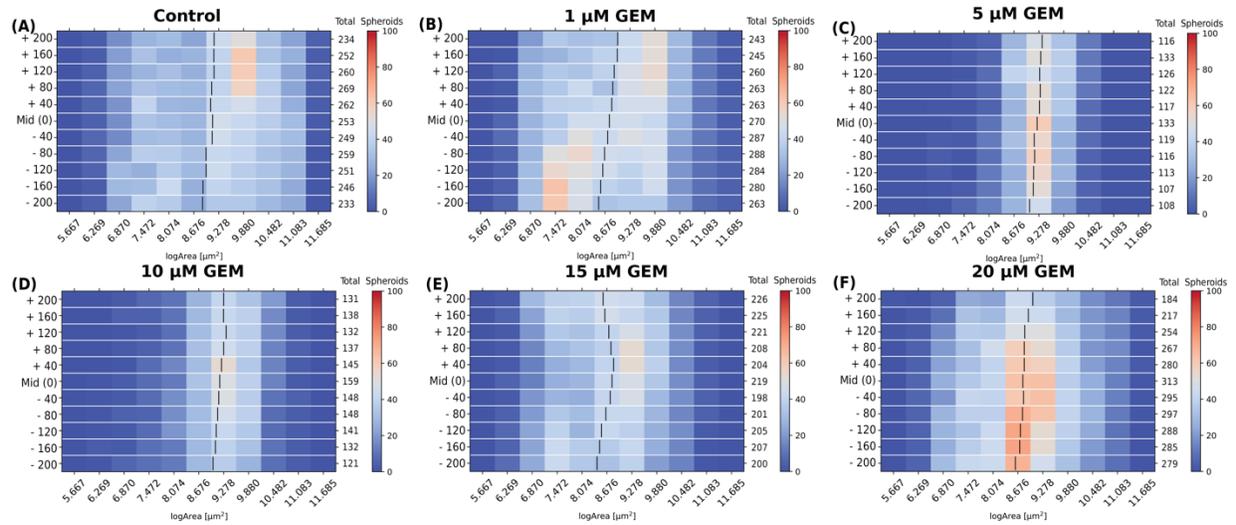

*Figure S30 - Day 6 2D histograms separated by the focal plane against the log(area) of all individually identified spheroids.*

### 4.9 Other Statistical Analyses

The Shapiro test was used to determine normality of all data before statistical hypothesis testing. Where data was significantly large for all groups being studied, the central limit theorem approximation of normality was used. All data is paired, so related t-test and repeated measures ANOVA with Tukey HSD post-hoc analyses were used for large data with two groups or more than two groups for comparison, respectively.

All analyses were implemented in Python using scipy, scipy.stats, pingouin, sklearn, math libraries, and the associated functions for the above calculations. The population stability index function came from a GitHub repository[3].

# 5 Experimental metadata

*Table S6: Image and Experimental hardware specifications.*

**Experimental/Sample**

| | |
|---|---|
| Experimenter Name | KJ Trettner |
| Experiment Description | imaging and metabolic activity monitoring of PDAC spheroids in the presence of increasing concentrations of nanoparticles |
| Experiment Date(s) | 2021-06-13 through 2021-06-20 , 2021-06-28 through 2021-07-04, 2022-04-27 through 2022-05-03, 2022-05-17 through 2022-05-23, 2023-04-05 through 2023-04-11 |
| Sample Description | pancreatic adenocarcinoma mouse line 8-14F-7: KRAS G12D, PTEN loss, COX2 overexpression, female, 2 weeks old at the time of sacrifice |
| Medium | Complete feeding medium, made as previously reported (1) |
| Temperature | 37C |
| $CO_2$ | 5% |

**Microscope hardware specifications**

| | |
|---|---|
| Microscope manufacturer(s) | ECHO, Perkin Elmer |
| Microscope model(s) | ECHO Revolve, Operetta CLS |
| Objective manufacturer ECHO | Olympus |
| Magnification/NA ECHO | 5x/0.16 |
| Objective manufacturer CLS | OperaPHX/OPRTCLS |
| Magnification/NA CLS | 5x/0.16 |
| Camera ECHO | 5 MP CMOS color camera |
| Camera CLS | 4.7Mpx sCMOS camera (2160x2160), 16-bit resolution, 6.5um pixel size |

**Image acquisition settings**

| | |
|---|---|
| Acquisition date(s) ECHO | 2021-06-13, 2021-06-17, 2021-06-19, 2021-06-28, 2021-07-02, 2021-07-04, 2022-04-27, 2022-04-01, 2022-05-03, 2022-05-19, 2022-05-23 |
| Acquisition date(s) CLS | 2023-04-05, 2023-04-06, 2023-04-07, 2023-04-08, 2023-04-09, 2023-04-10 |
| Illumination type ECHO | LED |
| Illumination type CLS | LED 740nm |
| Channel name | Brightfield |

*Table S7: List of datasets software provided. All files are located on Github.*

| File name | Description | Comments |
|---|---|---|
| longevity.csv | Operetta CLS 2D images analyzed by SAAVY | 320 µm midplane images only |
| mappings.csv | ECHO 2D images analyzed by SAAVY | |
| SAAVY-individual_spheroid-planar_longevity.xlsx | Operetta CLS 3D images analyzed by SAAVY | Include metadata |
| expert_color_match.xlsx | | Expert 1 (or 2) = the percent white value Expert 1 (or 2) assigned to a randomly generated greyscale image; Expert 1 and Expert 2 containing the same information. |
| SAAVYTorch-debug_1.1-longevity.xlsx | SAAVY analysis, metadata, and confusion matrix analysis | Include metadata from longevity images and confusion matrix classifications |
| SAAVYTorch-debug_1.1-mappings.xlsx | SAAVY analysis, metadata, and confusion matrix analysis | Include metadata from mappings images and confusion matrix classifications |
| SAAVY code | | https://github.com/armanilab/SAAVY |

*Table S8: List of terminology used in code and/or excel files*

| Variable or term | Description or definition |
|---|---|
| File | file name of the image analyzed |
| original_file | the original file name of the analyzed image (these were blinded before SAAVY analysis) |
| Image_Name | the filename that was assigned to the analyzed image during blinding |
| Count | total number of spheroids identified in the image |
| avg_viability | the averaged viability of all independently identified spheroid viabilities, measured from 0 to 100 |
| avg_circularity | a measure of how circular the spheroid is, where 1 is perfectly circular, averages from all independently identified spheroids |

| avg_intensity | the averaged intensity of all independently identified spheroids, measured from 0 to 255 for 8-bit color coding |
|---|---|
| pct_analyzed | the averaged percent of the image that was included as an identified spheroid region of interest |
| background_intensity | the averaged intensity of the background that was not included in an identified spheroid region of interest |
| avg_area | the averaged area of all independently identified spheroids |
| Truth | the actual percent white value of the randomly generate greyscale image |
| Expert 1 (or 2) | the percent white value Expert 1 (or 2) assigned to a randomly generated greyscale image in the expert_color_match data set |
| Day | the day of the experiment (beginning at Day 0 for initial seeding) on which the image was captured |
| gem_conc | the concentration of gemcitabine, a chemotherapeutic agent, added to the sample well |
| original_file | the original file name of the analyzed image (these were blinded before SAAVY analysis) |
| np_c | the concentration of nanoparticles doped into the sample, where appropriate for noisy background study |
| Expert1_viability | the viability (from 0 to 100%) that Expert 1 assigned to the image |
| Expert2_viability | the viability (from 0 to 100%) that Expert 2 assigned to the image |
| SAAVYTorch | the viability that SAAVY assigned to the analyzed image |
| CTG_viability | the viability (where appropriate matched data was available) that CTG 3D analysis output after normalization |
| Treatment? | yes/no if the image was a well sample that was treated with gemcitabine on Day 4 of the associated experiment |
| D0 ID | ground truth status for Day 0 spheroids, NaN if not Day 0 and 0 otherwise because insufficient time has passed to form spheroids |
| Expert 1 ID | based on the value that Expert 1 assigned to a sample when identifying the presence of spheroids. 0 for no spheroids and 1 for spheroids present |
| Expert 2 ID | based on the value that Expert 2 assigned to a sample when identifying the presence of spheroids. 0 for no spheroids and 1 for spheroids present |
| SAAVY ID | based on the value assigned to a sample based on SAAVY output, where no spheroids measured is a -1 output and otherwise is considered spheroid identified. 0 for no spheroids and 1 for spheroids present |
| CTG ID | the value assigned to a sample based on CTG output. |
| Expert 1 alive | based on the value that Expert 1 assigned to a sample when identifying if a spheroid was alive or dead. 0 for dead spheroids (viability of 0) and 1 for alive spheroids (spheroids with a viability greater than 0). |
| Expert 2 alive | based on the value that Expert 2 assigned to a sample when identifying if a spheroid was alive or dead. 0 for dead spheroids (viability of 0) and 1 for alive spheroids (spheroids with a viability greater than 0). |
| SAAVY alive | based on the value that SAAVY assigned to a sample when identifying if a spheroid was alive or dead. 0 for dead spheroids (viability of 0) and 1 for alive spheroids (spheroids with a viability greater than 0). |
| CTG alive | based on the value that CTG assigned to a sample when identifying if a spheroid was alive or dead. 0 for dead spheroids (viability of 0) and 1 for alive spheroids (spheroids with a viability greater than 0). |

# 6 SI References